\documentclass[conference]{IEEEtran}
\usepackage{cite}
\usepackage{amsmath,amssymb,amsfonts}
\usepackage{algorithmic}
\usepackage{graphicx}
\usepackage{textcomp}
\usepackage{xcolor}
\usepackage[hyphens]{url}

\usepackage[normalem]{ulem}

\usepackage{amsmath}
\usepackage{amssymb}
\usepackage{amsthm}
\usepackage{amsfonts}
\usepackage{graphicx} 
\graphicspath{ {./fig/} }
\usepackage{graphicx}
\usepackage{subfigure}
\usepackage{multirow}
\usepackage{fancyhdr}
\usepackage{xcolor}
\usepackage{physics}
\usepackage{cancel}
\usepackage[ruled,vlined,linesnumbered,resetcount,noend]{algorithm2e}
\usepackage{setspace}
\usepackage{todonotes}

\usepackage{syntax}
\usepackage{array}
\newcolumntype{?}{!{\vrule width 1pt}}

\usepackage[numbers,sort&compress]{natbib}

\usepackage{amsmath}
\usepackage{amssymb}
\usepackage{amsthm}
\usepackage{stmaryrd}

\newcommand{\myCompilerName}{Bosehedral} %will change to Bosehedral in camera ready version
\newcommand{\myCompilerNameSpace}{Bosehedral }

\providecommand{\TODO}[1]{{\protect\color{red}\noindent {\bf [TODO]}\emph{#1} {\bf [/TODO]}}}

\providecommand{\review}[1]{{\protect\color{green}\noindent {\bf[review]} 
{\bf [/review]}}}

\def\BibTeX{{\rm B\kern-.05em{\sc i\kern-.025em b}\kern-.08em
    T\kern-.1667em\lower.7ex\hbox{E}\kern-.125emX}}

\title{\LARGE \myCompilerName : Compiler Optimization for Bosonic Quantum Computing}
\author{$^1$Junyu Zhou, $^1$Yuhao Liu, $^2$Yunong Shi, $^3$Ali Javadi-Abhari, $^1$Gushu Li \\
$^1$University of Pennsylvania, $^2$AWS Quantum Technology, $^3$IBM Quantum}
\begin{document}
\maketitle
\thispagestyle{plain}
\pagestyle{plain}

%%%%%% -- PAPER CONTENT STARTS-- %%%%%%%%

\begin{abstract}

Bosonic quantum computing, based on the infinite-dimensional qumodes, has shown promise for various practical applications that are classically hard. However, the lack of compiler optimizations has hindered %\jz{
%utilizing 
its full potential. This paper introduces \myCompilerName, an efficient compiler optimization framework for (Gaussian) Boson sampling on Bosonic quantum hardware. \myCompilerName~overcomes the challenge of handling infinite-dimensional qumode gate matrices by performing all its program analysis %\jz{analyze ``whole'' program or any program analysis?} 
and optimizations at a higher algorithmic level, using a compact unitary matrix representation. It optimizes qumode gate decomposition and logical-to-physical qumode mapping, and introduces a tunable probabilistic gate dropout method. 
%to improve performance. 
Overall, \myCompilerName~significantly improves the performance by accurately approximating the original program with much fewer gates.
Our evaluation shows that \myCompilerNameSpace can largely reduce the program size but still maintain a high approximation fidelity, which can translate to significant end-to-end
application performance improvement.
%the effectiveness of \myCompilerName~in harnessing the potential of Bosonic quantum computing for practical applications.

\end{abstract}

\section{Introduction} \label{Introduction}

%\ys{The intro is a bit overwhelming, too many terms are introduced without context: continuous variable, harmonic oscillators, infinite dimensional qumodes, unitary. Suggest to start with the intuition and importance directly. }

%\ys{Example: Qubits and qumodes are two foundational models in the quantum mechanical world. Qubits,  embodies discrete quantum states, has been the mainstay of the quantum computing revolution; Qumodes, representing continuous-variable quantum systems, or Bosonic  systems, are rapidly gaining traction, offering a compelling suite of advantages for quantum computation. These include the prolonged coherence times characteristic of qumodes in superconducting cavities, the seamless transduction between stationary and flying quantum information, and an inherent efficiency in encoding specific computational tasks. }

%\ys{Suggest to add 1 figure to visually compare qubits and qumodes as well as showcase the computational complexity of compiling qumodes}

\begin{figure}[t]
    \centering
    \includegraphics[width=0.8\linewidth]{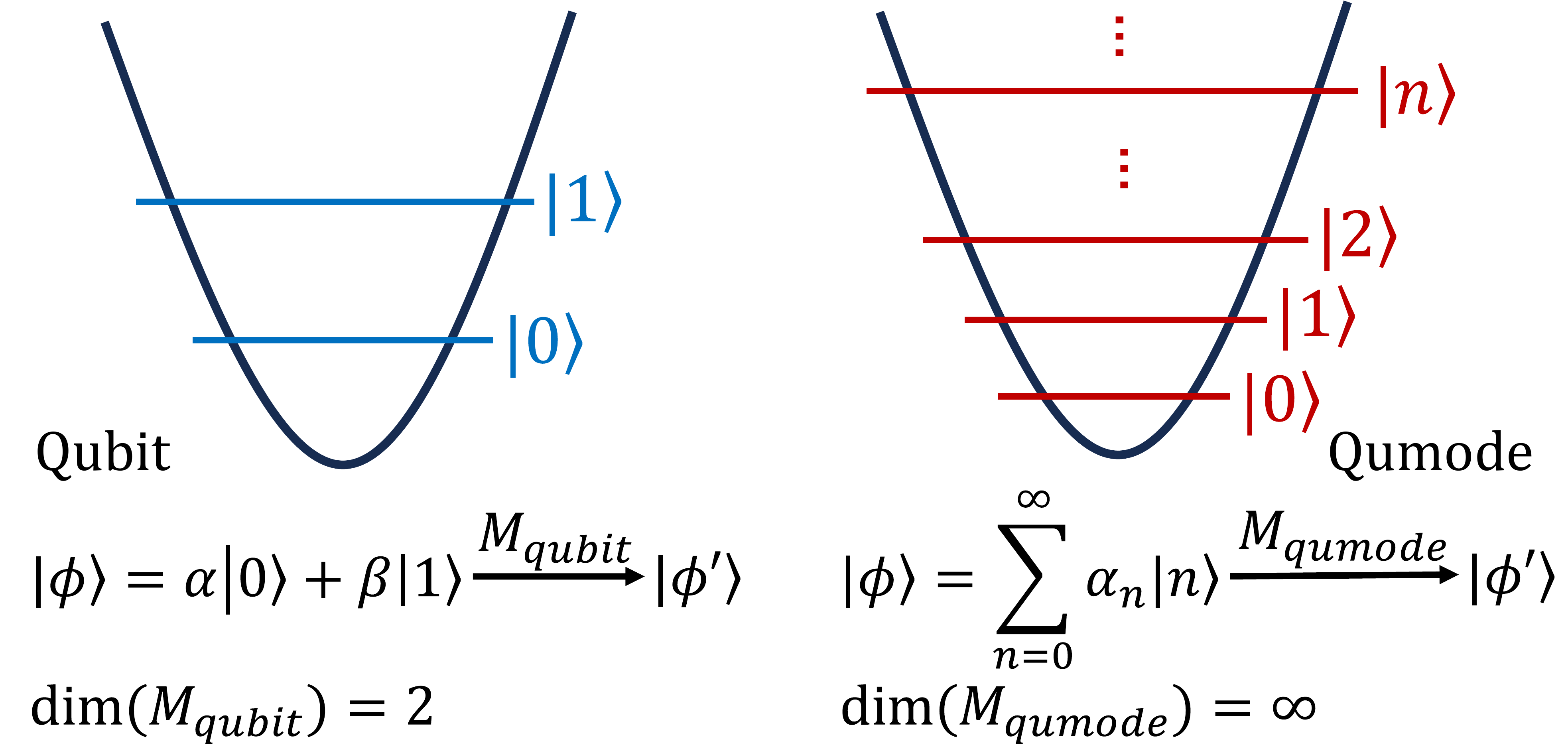}
    \vspace{-5pt}
    \caption{Qubit vs Qumode}
    \vspace{-20pt}
    \label{fig:qubitqumode}
\end{figure}

Bosonic quantum computing, also known as continuous-variable quantum computing~\cite{lloyd1999quantum},  is built upon Bosonic modes.
% ~\ali{In English, words that are named after a person are always capitalized (e.g. Hamiltonian, Hermitian). So Boson and Bosonic should always be capital-B. I fixed a few but there are many throughout the paper.}
Contrary to qubit-based discrete variable quantum
computing, the basic information processing unit in Bosonic quantum computing, the qumode, %is by itself a
%quantum harmonic oscillator and 
by itself
has an infinite-dimensional state space (shown in Fig.~\ref{fig:qubitqumode}). 
Bosonic quantum computing (QC) is attractive for multiple reasons: long lifetimes of qumodes (e.g., superconducting cavities~\cite{wang2020efficient}), the ability to transduce between stationary and flying (e.g., photonic) information~\cite{narla2016robust}, etc. 
In particular, Bosonic QC has strong built-in information encoding and processing capability to naturally and efficiently encode certain computations, such as Boson sampling~\cite{aaronson2011the} and Gaussian Boson Sampling~\cite{hamilton2017gaussian}, with various practical applications~\cite{bromley2020applications} (e.g., graph clique~\cite{banchi2020molecular}, graph similarity~\cite{schuld2019quantum}, point process~\cite{jahangiri2020point}, and molecule vibrational spectra simulation~\cite{huh2015boson}) that are hard for classical computing. 
Recently, the quantum advantage~\cite{zhong2020quantum,zhong2021phase,madsen2022quantum} has been experimentally demonstrated on several Bosonic QC platforms.
Several startups are pursuing the commercialization of Bosonic QC with various technologies~\cite{killoran2019strawberry,maring2023general,qci,Taballione2023modeuniversal}.

%with strong built-in information encoding and processing capability.

%The computational paradigms in Bosonic quantum computing, such as Boson sampling~\cite{aaronson2011the} and Gaussian Boson sampling~\cite{hamilton2017gaussian}, can naturally encode  
%Boson sampling is one common computation paradigm on bosonic quantum devices and has 
%diverse practical applications~\cite{bromley2020applications} (e.g., graph clique~\cite{banchi2020molecular}, graph similarity~\cite{schuld2019quantum}, point process~\cite{jahangiri2020point}, and molecule vibrational spectra simulation~\cite{huh2015boson}), which are usually computationally hard on classical computers. Bosonic quantum computing can be implemented with various technologies, such as superconducting circuits~\cite{girvin2014circuit,wang2020efficient}, quantum photonics~\cite{arrazola2021quantum,taballione202220}, and trapped ions~\cite{ortiz2017continuous, chen2021quantum}.
%Recent works have demonstrated quantum computational advantage~\cite{zhong2020quantum,zhong2021phase,madsen2022quantum} and early error correction capabilities~\cite{sivak2023realtime} using Bosonic quantum hardware platforms. \ali{Bosonic quantum computing is attractive for multiple reasons: long lifetimes of qumodes (e.g. superconducting cavities), the ability to transduce between stationary and flying (e.g. photonic) information, and the ability to naturally and efficiently encode certain computations. Several startups pursue the commercialization of these platforms~\cite{killoran2019strawberry,maring2023general,qci}} 

Despite the great hardware progress, the development of software and compiler optimizations for Bosonic QC is far behind. 
Early efforts on programming and compilation for Bosonic QC, including Strawberry Fields~\cite{killoran2019strawberry}, Bosonic Qiskit~\cite{stavenger2022c2qa}, and Perceval~\cite{maring2023general}, provide basic programming interfaces and operation decomposition functions~\cite{reck1994experimental,clements2016optimal} but miss program optimizations. 
%~\ali{the following detailed sentences fit better in related work section.}\cite{reck1994experimental} and \cite{clements2016optimal} studied the basic approaches to decompose linear interferometers into single- and two-qumode gates with a fixed cost. %, which are the basic approaches to implement  
%\cite{kang2023leveraging} focused on low-level pulse compilation for specific qumode operations with analytical solutions for a single superconducting qubit-qumode pair. 
%\cite{rueda2021continuous} analyzed continuous variable quantum compilation but failed to provide actual optimizations. 
To summarize, compilation for Bosonic QC is in its infancy. In contrast to qubit-based QC, rich libraries of compilation passes~\cite{Qiskit,sivarajah2021tket} are non-existent, hindering the full exploitation of these computing platforms. %\ys{need to answer why now and why can't we wait.}

%The challenge of desving compiler optimizations for bosonic quantum computing comes from each individual qumode having an infinite-dimensional state space
The complexity of devising compiler optimizations for Bosonic QC arises from the inherent infinite-dimensional state space of each individual qumode~\cite{lloyd1999quantum}.
All the gates manipulating the state of qumodes, even for a single qumode, have infinite-dimensional gate matrices. 
Thus, it is highly non-trivial for the quantum compiler that runs on a classical computer to derive equivalent program transformations for optimization.
Traditional techniques employed for quantum program transformations on qubit-based devices, which rely on gate matrices, encounter substantial obstacles when confronting the infinite-dimensional qumode gate matrices.
%is hard to be directly extended to qumodes as all qumode gates have infinite-dimensional gate matrices.

%As the first attempt to successfully overcomes this grand challenge the best, this paper 

%\ys{it will be much easier for reviewers if we have another fig here (or a subfig in fig 1 suggested above), the first fig we have now is a circuit diagram which has no differences to qubit based circuits to reviewers}
In this paper, we tackle this grand challenge by investigating the compiler optimization in the \textbf{high-level semantics} of Bosonic QC.
Instead of directly handling the infinite-dimensional gate matrices applied on the qumode state, the unique high-level semantics of Bosonic QC allow us to analyze and optimize the program components efficiently and effectively in a compact data structure.
In particular, we focus on the linear interferometer, the pivotal component containing most of the gates in a (Gaussian) Boson sampling program (a widely used Bosonic QC paradigm illustrated in Fig.~\ref{fig:gaussianbosonsampling}). A linear interferometer can be considered as a unitary matrix of size $N\times N$ for an $N$-qumode program without losing any algorithmic information. 
All the program transformation and optimization techniques proposed in this paper can be reasoned about in the high-level unitary matrix.

To this end, we propose \myCompilerName, an effective and efficient compiler optimization framework for (Gaussian) Boson sampling on Bosonic quantum hardware.
In contrast to peep-hole approximations performed by qubit-based compilers~\cite{Qiskit,sivarajah2021tket}, \myCompilerNameSpace can approximate and simplify circuits at a large scale by exploiting the global program semantics.
%all the program analysis and compilation optimizations are performed over the high-level representation of the linear interferometer, a unitary matrix of size $N\times N$ for an $N$-qumode program without losing any algorithmic information. 
%formidable challenge 
%posed by the unique characteristics of Bosonic quantum computing
%and
%propose \myCompilerName, an effective and efficient compiler optimization framework for (Gaussian) Boson sampling on Bosonic quantum hardware.
%\textbf{First}, \myCompilerName~studies the compilation of the linear interferometer, the pivotal component containing most of the gates in a (Gaussian) Boson sampling program (illustrated in Figure~\ref{fig:gaussianbosonsampling}), at a higher algorithmic level. 
%Instead of an infinite-dimensional gate matrix applied on the qumode state, all the program analysis and compilation optimizations are performed over the high-level representation of the linear interferometer, a unitary matrix of size $N\times N$ for an $N$-qumode program without losing any algorithmic information. 
%This ensures that the follow-up compilation process can be efficient and effective.
%This facilitates an efficient and effective subsequent compilation process.
\textbf{First}, \myCompilerNameSpace optimizes the \textit{qumode gate decomposition} for linear interferometers, which can be considered as %performing matrix elimination in the high-level unitary. 
%The gate decomposition for linear interferometers can be considered as performing matrix elimination in the high-level unitary. 
%\myCompilerNameSpace 
adjusting the elimination patterns for different input high-level unitary matrices to maximize the occurrence of two-qumode gates with very small rotation angles.
These small-angle gates are functionally akin to the identity and can be safely disregarded to reduce the gate error with minimal effect on the overall program semantics.
\textbf{Second}, \myCompilerNameSpace modifies the \textit{logical-to-physical qumode mapping}, which can be considered as applying row and column permutations on the high-level unitary.
Bose can find better qumode mapping to further reduce rotation angles in the compiled two-qumode gates. The remapping is implemented via re-labeling the physical qumodes without any execution overhead.
%\myCompilerName~offers optimization for two essential tasks, \textit{qumode gate decomposition} and \textit{logical-to-physical qumode mapping}.
%Gate decomposition is optimized by adjusting the elimination patterns in the high-level unitary matrix to maximize the occurrence of two-qumode gates with very small rotation angles.
%These small-angle gates are functionally akin to the identity and can be safely disregarded to reduce the gate error with minimal effect on the overall program semantics.
%Through qumode mapping optimization we can further reduce rotation angles in the compiled two-qumode gates by just re-labeling the physical qumodes without any execution overhead.
\textbf{Third}, \myCompilerName~comes with a \textit{tunable probabilistic gate dropout} method to approximate linear interferometers.
%\textcolor{red}{
Gates with exceedingly small rotation angles will very likely be dropped, while gates with rotation angles near the threshold will be dropped probabilistically to average over the algorithmic approximation errors. %}
As a result, \myCompilerName~can approximate (Gaussian) Boson sampling accurately using considerably fewer gates and thus substantially enhances the overall performance by largely mitigating the hardware error effects.
%\textbf{}

\textbf{Overall approximation effect reasoning}
One key advantage of \myCompilerNameSpace is that it can easily reason about the overall effect of the approximation during compilation time.
After the gate decomposition and dropout, the global high-level semantics of the approximated program can be reconstructed by reversing the elimination process on the high-level unitary matrix.
This is hard to achieve in the qubit-based approximation compilations because they usually rely on the unscalable low-level gate semantics and the gate matrices are exponentially large as the number of qubits increases.

Our experimental results show that \myCompilerName~can reduce gate by $\sim25\%$ to $40\%$ but maintain program fidelity of $\sim98\%$ to $99.9\%$ for various benchmarks and underlying architectures.
Compared with the baseline Strawberry Fields~\cite{killoran2019strawberry}, the divergence %(Jensen-Shannon Divergence~\cite{Jensen–S53:online}) 
between the sampled output distribution in noisy simulation and the ideal %standard \ali{ideal?}
output distribution is reduced by $26.1\%$ on average, which translates to significant end-to-end application performance improvement as demonstrated in our detailed application studies. 
%Our end-to-end studies also confirm that \myCompilerName~can significantly increase the end-to-end success probability of graph sampling and refle
%\TODO{to be completed} 
%\textcolor{blue}{We have done experiments on different kinds of benchmarks: dense subgraph, maximum clique, graph similarity, and vibration molecule as well as different device architectures: 6-by6, 5-by-7, and 3-by-8 device. In each problem and architecture, \myCompilerName~has a significant effect on reducing the total Beamsplitter gates. On average, we have maintained 99.98\%, 99.99\%, 99.90\%, and 98.00\% accuracy for unitary approximation while cutting 29.8\%, 25.5\%, 26.5\%, and 40.7\% BS gates for four benchmarks respectively. The Jensen-Shannon Divergence to original non-loss samples are reduced by xx\%, xx\%, xx\%, and xx\%. In addition, we also show the results under the end-to-end metric, we increase the probability that successfully finding dense subgraph and clique by xx\% and xx\% respectively, reducing the distance of graph feature vectors to non-loss one by xx\%, and our method can confront the problem that Spectra shifting to zero when encounter the loss for vibration molecule.}

Our major contributions can be summarized as follows:
\begin{enumerate}
    \item We proposed \myCompilerName, the first efficient and effective compiler optimization framework for (Gaussian) Boson sampling in Bosonic quantum computing.
    \item \myCompilerName~overcomes the challenges of infinite-dimensional gate matrices by performing program analysis and compiler optimization at a high-level representation. %\ali{In contrast to peep-hole approximations performed by qubit-based compilers, we are able to approximate and simplify circuits at a large scale by exploiting global program semantics.}% of (Gaussian) boson sampling.
    %\ali{i'm not sure i understand this. in practice people truncate dimensions with tiny loss. also in qubit compilers we rarely track exponentially-large gate matrices. i think the novelty here is in how we can reason about approximations during decomposition+mapping.}
    \item We proposed several compiler optimization algorithms for qumode gate decomposition, logical-to-physical qumode mapping, and probabilistic gate dropout for program simplification.
    \item Our evaluation shows that \myCompilerName~can outperforms baseline Bosonic quantum compilers by significantly improving the execution fidelity and the end-to-end application performance for various benchmarks and architectures.
\end{enumerate}

\section{Background} \label{Background}

This section introduces the necessary background to understand the proposed compiler optimization techniques. We recommend~\cite{lloyd1999quantum,weedbrook2012gaussian, hamilton2017gaussian} for more details about general Bosonic QC and Gaussian Boson sampling.
%we first provide the basic knowledge of continuous variable quantum information, see~\cite{weedbrook2012gaussian} for more details. Then, we introduce the framework of Gaussian Boson Sampling~\cite{hamilton2017gaussian}, which can be viewed as an algorithm that demonstrates quantum advantage.

\subsection{Continuous Variable Quantum Computing} \label{gate introduce}
%\textcolor{blue}{In discrete quantum computation and quantum information, the basic information processing unit is a qubit. The state of qubit can be described by a superposition of two basic states: $\vert 0 \rangle$ and $\vert 1 \rangle$. In the continuous variable scenario, the basic bit is bosonic mode or qumode. Its state lies in infinite-dimensional Hilbert space spanned by Fock basis: $\{\vert n \rangle\}_{n=0}^{\infty}$.
%For the $i$-th qumode, we have the annihilation operator $\hat{a}_{i}$ and the creation operator $\hat{a}^{\dagger}_{i}$ defined as follows:}
In discrete variable quantum computing, the basic information processing unit is a qubit, whose state lies in a two-dimensional Hilbert space spanned by two basis states $\ket{0}$ and $\ket{1}$. %The state of qubit can be described by a superposition of two basic states: $\vert 0 \rangle$ and $\vert 1 \rangle$.  
In contrast, in the continuous variable scenario, the basic information processing unit is a Bosonic mode or qumode. Its state lies in infinite-dimensional Hilbert space spanned by the Fock basis: $\{\vert n \rangle\}_{n=0}^{\infty}$.
For the $k$-th qumode, the annihilation operator $\hat{a}_{k}$ and the creation operator $\hat{a}^{\dagger}_{k}$ are defined as follows:
%This infinite-dimensional Hilbert space is spanned by Fock basis: $\{\vert n \rangle\}_{n=0}^{\infty}$. To describe the one-mode system, such as a harmonic oscillator, we need the annihilation operator $\hat{a}$ and the creation operator $\hat{a}^{\dagger}$, which have the following properties:
\begin{equation*}
   \begin{array}{ll}
   \hat{a}_{k} \vert n \rangle_{k}  = \sqrt{n}\vert n-1 \rangle_{k} \enspace &\text{for} \enspace n \geq 1\\
   \hat{a}^{\dagger}_{k} \vert n \rangle_{k}= \sqrt{n+1}\vert n+1 \rangle_{k} \enspace &\text{for} \enspace n \geq 0
   \end{array}
\end{equation*}
and $\hat{a}_{k} \vert 0 \rangle = 0$. 
%Note that all the $\hat{a}_{i}$'s and  $\hat{a}^{\dagger}_{i}$'s are infinite-dimensional.
For an $N$-qumode system, we have the following operator vectors:
\begin{equation*}
   \begin{array}{ll}
   \hat{a} = (\hat{a}_{1}, \hat{a}_{2}, \ldots, \hat{a}_{N})^\intercal, \
   \hat{a}^{\dagger} = (\hat{a}_{1}^{\dagger}, \hat{a}_{2}^{\dagger}, \ldots, \hat{a}_{N}^{\dagger})^\intercal
   \end{array}
\end{equation*}
%A linear transformation of N modes bosonic system can be written as:
%\begin{equation*}
%   \begin{array}{ll}
%   \hat{a} \rightarrow \textbf{A} \hat{a} + \textbf{b}
%   \end{array}
%\end{equation*}
%where $\textbf{A}$ is an N by N matrix and $\textbf{b}$ is an N-dimensional vector.

% Next, we place the basic gates and their circuit representation we will use in this paper.
%Some common qumode gates used in this paper %can then be defined with the operator vectors.
Qumode gates are usually defined with the $\hat{a}_{k}$'s and  $\hat{a}^{\dagger}_{k}$'s.
Some common qumode gates used in this paper are listed in Fig.~\ref{fig:gaussianbosonsampling} and introduced in the following. \textbf{Squeezing Gate} is a single-qumode gate denoted by `$S$'. A squeezing gate applied on the $k$-th qumode is defined as: $S(\alpha)=exp\left(\frac{1}{2}(\alpha^* \hat{a}_{k}^2-\alpha \hat{a}_{k}^{\dagger2})\right)$, where $\alpha \in \mathbb{C}$.
%: See the boxes containing $S$ in Figure~\ref{fig:gaussianbosonsampling} for the gate, the one mode squeezing gate is given by:
%\begin{equation*}
%   \begin{array}{ll}
%   S(re^{i\phi})=exp\left(\frac{1}{2}(re^{-i\phi}\hat{a}_{i}^2-re^{i\phi}\hat{a}_{i}^{\dagger2})\right)
%   \end{array}
%\end{equation*}
%the transformation of annihilation operator is:
%\begin{equation*}
%   \begin{array}{ll}
%   S(re^{i\phi})^{\dagger}\hat{a}S(re^{i\phi}) = \hat{a}\cosh{r}-\hat{a}^{\dagger}e^{i\phi}\sinh{r}
%   \end{array}
%\end{equation*}
\textbf{Phase Shifter} is a single-qumode gate denoted by `$R$'. A phase shifter applied on the $k$-th qumode is defined as: $R(\phi)=exp(i\phi\hat{a}_{k}^{\dagger}\hat{a}_{k})$, where $\phi \in \mathbb{R}$.
%: See the boxes containing $R$ in Figure~\ref{fig:gaussianbosonsampling} for the gate, the one mode phase shifter is given by:
%\begin{equation*}
%   \begin{array}{ll}
%   R(\phi)=exp(i\phi\hat{a}_{i}^{\dagger}\hat{a}_{i})
%   \end{array}
%\end{equation*}
%the transformation of annihilation operator is:
%\begin{equation*}
%   \begin{array}{ll}
%   R(\phi)^{\dagger}\hat{a}R(\phi) = \hat{a}e^{i\phi}
%   \end{array}
%\end{equation*}
\textbf{Beamsplitter} is a two-qumode gate denoted by `$BS$'.  A Beamsplitter applied on the $k$-th and $l$-th qumodes is defined as: $BS(\theta,\phi)=exp\left(\theta(e^{i\phi}\hat{a}_{k}\hat{a}_{l}^{\dagger}-e^{-i\phi}\hat{a}_{k}^{\dagger}\hat{a}_{l})\right)$, where $\phi $ and $\theta \in \mathbb{R}$.
%: See the boxes containing $BS$ in Figure~\ref{fig:gaussianbosonsampling} for the gate, the two mode Beamsplitter is given by:
%\begin{equation*}
%   \begin{array}{ll}
%   B(\theta,\phi)=exp\left(\theta(e^{i\phi}\hat{a}_{i}\hat{a}_{j}^{\dagger}-e^{-i\phi}\hat{a}_{i}^{\dagger}\hat{a}_{j})\right)
%   \end{array}
%\end{equation*}
%the transformation of annihilation operator is:
%\begin{equation*}
%   \begin{array}{ll}
%   B(\theta,\phi)^{\dagger}\hat{a}_{1}B(\theta,\phi) = \hat{a}_{1}\cos{\theta}-\hat{a}_{2}e^{-i\phi}\sin{\theta}\\
%   B(\theta,\phi)^{\dagger}\hat{a}_{2}B(\theta,\phi) = \hat{a}_{1}e^{i\phi}\sin{\theta}+\hat{a}_{2}\cos{\theta}
%   \end{array}
%\end{equation*}
\textbf{Displacement} is a single-qumode gate denoted by `$D$'. A displacement gate applied on the $k$-th qumode is defined as:  
$ D(\alpha)=exp(\alpha\hat{a}_{k}^{\dagger}-\alpha^*\hat{a}_{k})$, where $\alpha \in \mathbb{C}$.
%: See the boxes containing $D$ in Figure~\ref{fig:gaussianbosonsampling} for the gate, the one mode displacement is given by:
%\begin{equation*}
%  \begin{array}{ll}
%  D(\alpha)=exp(\alpha\hat{a}^{\dagger}-\alpha^*\hat{a})
%  \end{array}
%\end{equation*}
%the transformation of annihilation operator is:
%\begin{equation*}
%   \begin{array}{ll}
%   D(\alpha)^{\dagger}\hat{a}D(\alpha) = \hat{a}+\alpha
%   \end{array}
%\end{equation*}
%\begin{figure}[t]
%    \centering
%    \includegraphics[width=1.0\columnwidth]{fig/basic gate.jpg}
%    \vspace{-15pt}
%    \caption{\textbf{Basic Gate}}
%    \vspace{-5pt}
%    \label{fig:basicgate}
%\end{figure}
%\textcolor{blue}{
Note that all these gates have \textit{infinitely-large gate matrices} as all $\hat{a}_{k}$s and $\hat{a}_{k}^{\dagger}$s are infinite-dimensional.%}

\subsection{Gaussian Boson Sampling}\label{Gaussian Boson Sampling}

\begin{figure}[t]
    \centering
    \includegraphics[width=0.8\linewidth]{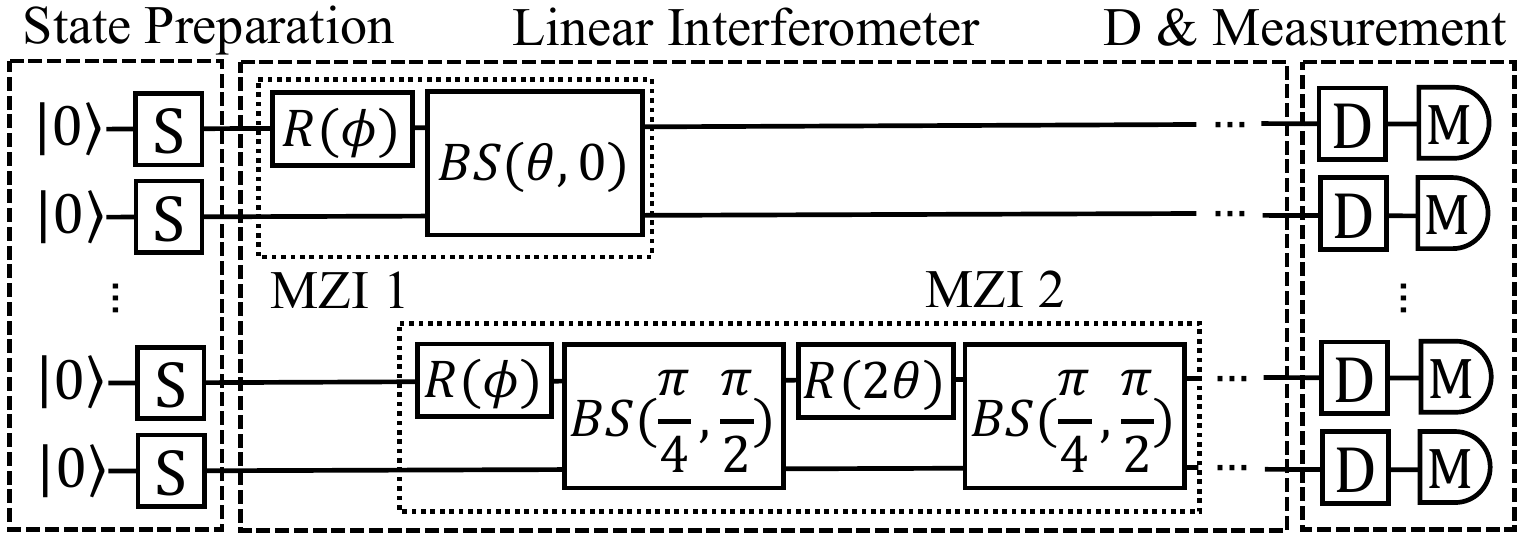}
    \vspace{-10pt}
    \caption{Overview of a Gaussian Boson sampling (GBS) program}
    \vspace{-15pt}
    \label{fig:gaussianbosonsampling}
\end{figure}
%\TODO{Say that our technique can be applied to general BS. We will use GBS to introduce the techniques in this paper}
%In this paper, we will illustrate our proposed compiler optimizations using Gaussian Boson sampling (GBS), but these proposed optimizations can be directly applied to general Boson sampling without modification since their differences are mostly in state preparation.
Gaussian Boson sampling (GBS) is a commonly used computation paradigm in Bosonic QC. 
As depicted in Fig.~\ref{fig:gaussianbosonsampling}, a GBS program typically has three major steps: state preparation, linear interferometer, and measurement.
Usually, all qumodes are initialized to the vacuum state $\ket{0}$ with photon count $0$. %\ali{
(Note that in Bosonic QC literature, many terms such as photons, interferometers, etc. are shared with photonics. However, in this paper, they represent purely software concepts, and % as we described, this technology 
can be implemented via diverse hardware platforms.)
%}
State preparation adds some photons to the system to prepare an initial state, which is done by applying squeezing gates in GBS. 
The prepared state will then be fed into a linear interferometer to perform a special calculation where the \textbf{high-level semantics} is a unitary matrix $\mathbf{U}$ applied on the vector of annihilation operators of the system and denoted as:
\begin{equation*}
   \begin{array}{ll}
   \hat{a} \rightarrow \hat{a}' = \mathbf{U} \hat{a}, \ \text{where} \  \hat{a} = (\hat{a}_{1}, \hat{a}_{2}, \ldots, \hat{a}_{N})^\intercal.
   \end{array}
\end{equation*}
%\textcolor{blue}{
%and $\hat{a'} = (\hat{a'}_{1}, \hat{a'}_{2}, \ldots, \hat{a'}_{N})$ are view as column vectors, see \cite{bromley2020applications} for more details.}
%\TODO{add reference for this}
Finally, qumodes are measured. These measurements are typically performed on the Fock basis for GBS programs. Sometimes displacement gates are applied before the measurement, depending on the application. 
%is a process that prepares Gaussian states and measures them in the Fock basis, which will give us output sample $\mathbf{S}=\left(s_1,s_2,\ldots,s_N\right)$, here $s_i$ stands for the photon number we detect at mode i. 

%A pure Gaussian state can be prepared by the circuit in Figure~\ref{fig:gaussianbosonsampling}, in which Squeezing, linear hhinterferometry, and Displacement are used. 

%The linear interferometry can be described by a unitary matrix U. The transformation of the annihilation operator can be written as:
%\begin{equation*}
%   \begin{array}{ll}
%   \hat{a} \rightarrow \hat{a}' = \textbf{U} \hat{a} 
%   \end{array}
%\end{equation*}
%where $\hat{a} = (\hat{a}_{1}, \hat{a}_{2}, \ldots, \hat{a}_{N})$. \review{Why directly comes to the annihilation operator as $\hat{a}\rightarrow\hat{a}'=\vb{U}\hat{a}$? I mean what is the connection between linear interferometry with a single annihilation operator?}
%\TODO{add reference for this}

The linear interferometer is the largest building block in GBS and needs to be decomposed into basic single- and two-qumode gates to be executed on hardware.
A unitary matrix $\mathbf{U}$ has the following decomposition formula \cite{reck1994experimental}:
\begin{equation}\footnotesize
   \mathbf{U} = \Lambda\left(\prod \limits_{\left(m,n\right) \in \mathcal{S}} \mathbf{T}_{m,n}\left(\theta,\phi\right)\right)
   \label{eq1}
\end{equation}
Here $\mathbf{T}_{m,n}(\theta,\phi)$ is a rotation in two-dimensional subspace, and $\mathcal{S}$ is a set containing pairs that describe the information of the dimension where the rotation is acting on. We also call the parameter $\theta$ as the rotation angle. 
$\Lambda$ is the diagonal matrix whose diagonal entries $\lambda_{ii}$ have modules $\abs{\lambda_{ii}}=1$. The matrix representation of $\footnotesize \mathbf{T}_{m,n}(\theta,\phi)$ is:
\begin{equation*}\footnotesize
    \mathbf{T}_{m,n}\left(\theta,\phi\right)=
    \bordermatrix{
          &            & \text{column}\ m &        & \text{column}\ n &   \cr
          & \mathbb{I} & 0 & \cdots & 0 & 0 \cr
       \text{row}\ m & 0 & e^{i\phi}\cos\left(\theta\right) & \cdots & -\sin\left(\theta\right) & 0 \cr
          & \vdots & \vdots & \mathbb{I} & \vdots & \vdots \cr
       \text{row}\ n & 0 & e^{i\phi}\sin\left(\theta\right) & \cdots & \cos\left(\theta\right) & 0 \cr
          & 0 & 0 & \cdots & 0 & \mathbb{I}
    }
\end{equation*}
Each  $\mathbf{T}_{m,n}\left(\theta,\phi\right)$ can be implemented by an Mach-Zehnder interferometer (MZI)~\cite{rarity1990two} circuit block applied on the $m$-th and $n$-th qumode~\cite{reck1994experimental}. %\ali{This is the most expensive part of the computation due to noise.  (cite some recent experimental numbers here to say MZI is 10x (?) more faulty than others?)}
This is the most expensive part of the computation due to noise.
The error rate of a Beamsplitter can be over $10\times$ higher than that of other single-qumode gates~\cite{chapman2023high,wang2020efficient}.
%\jz{Beamsplitter around 99\% fidelity~\cite{chang2021low}}
One MZI block can be realized using a Phase Shifter $R(\phi)$ on qumode $m$ and a Beamsplitter $BS(\theta, 0)$ on qumodes $m$ and $n$ (`MZI 1' in Fig.~\ref{fig:gaussianbosonsampling}).
Some Bosonic quantum hardware platforms only natively support fixed $50:50$ Beamsplitters $BS(\pi /4, \pi/2)$~\cite{Carolan2015Universal}. In this way, one MZI block can be implemented with two Phase Shifters and two Beamsplitters (`MZI 2' in Fig.~\ref{fig:gaussianbosonsampling}).
The discussion in this paper assumes the first implementation, but our proposed techniques are equally effective if using the second implementation.
%See the boxes containing $R$ and $BS$ in Figure~\ref{fig:gaussianbosonsampling} for the representation of gates. This combination is also called the MZI block. \textcolor{blue}{MZI block has two parameters as well as two representations, there are shown in \ref{fig:gaussianbosonsampling}. The first representation has one Phase Shifter and one Beamsplitter, they contain one parameter respectively, the parameter for $BS$ is $\theta$, and parameter $\phi$ in $BS$ gate are always zero in the MZI block. The second representation has two Phase Shifters and two $50:50$ Beamsplitters, the two Phase Shifters contain one parameter respectively, and the $\theta$ for two $BS$ gates are set to be $\pi /4$ in this case.}

% and another is for Beamsplitter. The $\phi$ of Beamsplitter in MZI is always $0$, which means using a $50:50$ Beamsplitter.   

\iffalse
Since this decomposition process is not unique, the two decomposition methods have different circuit shapes exist. \cite{reck1994experimental} introduced a circuit of triangular form to implement the unitary, and the circuit in \cite{clements2016optimal} takes a rectangular form. For an $N$ modes system, both of their circuits need $N(N-1)/2$ numbers of the MZI blocks, and their framework is fixed. 

\fi
%\TODO{needs recheck}

% \begin{figure}[t]
%     \centering
%     \includegraphics[width=1.0\columnwidth]{fig/decomposition structure.png}
%     \vspace{-15pt}
%     \caption{\textbf{Decomposition Structure}}
%     \vspace{-5pt}
%     \label{fig:decompositionstructure}
% \end{figure}

\subsection{Application Encoding in GBS}

%We place a brief introduction to the application of GBS device, for more details please see~\cite{bromley2020applications}.

%\textbf{Graph Problem}:

%It has been shown that for a Gaussian state with zero means, which can be prepared by using only squeezing and linear interferometry, the probability $Pr\left(S\right)$ of observing an output $\mathbf{S}=\left(s_1,s_2,\ldots,s_N\right)$ is given by
%\begin{equation*}
%   Pr\left(\mathbf{S}\right) = \frac{1}{\sqrt{\mathbf{det\left(Q\right)}}} \frac{ Haf\left(\mathbf{A} _S\right) }{s_1!s_2!\ldots s_N!} 
%\end{equation*}
%here Haf is a matrix function called hafnian, calculating the hafnian is a $\sharp$P-hard problem, which means it can not efficiently be computed on classical computer. 

%Matrices $\mathbf{Q}$ and $\mathbf{A}$ are completely determined by the covariance matrix $\Sigma$ of the Gaussian state. $\mathbf{A} _S$ is a submatrix of $\mathbf{A}$ by repeating columns and rows $(i, i+m)$ of $\mathbf{A}$ $s_i$ times. 
%\begin{equation*}
%\begin{array}{ll}
%     &\mathbf{Q} = \Sigma + \mathbb{I}_{N}/2  \\[7pt]
%     &\mathbf{A} = \mathbf{X} \left( \mathbb{I}_{N}-\mathbf{Q}^{-1}\right) \\[7pt]
%     &\mathbf{X} = \begin{pmatrix}
%     0 & \mathbb{I}_{N}\\
%     \mathbb{I}_{N} & 0 
%    \end{pmatrix} 
%\end{array}
%\end{equation*}

%\TODO{Just say that these applications can be programmed by changing the parameters in sequeezing/MZI}

%GBS has famously been proposed as a suitable candidate for demonstrating quantum computational advantage in the near term~\cite{zhong2020quantum,zhong2021phase,madsen2022quantum}. However, mM
Many important applications can naturally be mapped to GBS, e.g., graph clique~\cite{banchi2020molecular}, graph similarity~\cite{schuld2019quantum}, point process~\cite{jahangiri2020point}, and molecule vibrational spectra simulation~\cite{huh2015boson}, and then solved with Bosonic QC~\cite{bromley2020applications}.
These applications are hard on classical computers and are even not efficient in qubit-based QC. 
For example, molecule vibrational spectra simulation is modeling Bosons which have infinite-dimension state space. Bosonic QC has natural advantages here as the Bosons can be directly encoded in qumodes and the Boson operations are also usually natively supported on Bosonic QC hardware. 
These applications of different purposes can be programmed onto GBS devices by changing the qumode gates' parameters and the linear interferometer's unitary matrix.
The three-step overall GBS program structure remains unchanged.
How to turn the application into its corresponding program parameter setup can be found in~\cite{bromley2020applications}.
The techniques in this paper do not rely on any application-specific information. % and only need the high-level semantics of the linear interferometer, the unitary $\mathbf{U}$.

\section{Problem Formulation} \label{Problem Formulation}

% \begin{figure}[t]
%     \centering
%     \includegraphics[width=1.0\columnwidth]{fig/device structure.png}
%     \vspace{-15pt}
%     \caption{\textbf{Device Structure}}
%     \vspace{-5pt}
%     \label{fig:devicestructure}
% \end{figure}

%\textcolor{blue}{
%A main process in the Gaussian Boson Sampling is implementing linear interferometry. Our optimization process sheds light on efficiently using Bosonic quantum gates to construct this linear transformation.} 

%The techniques in this paper focus on optimizing the implementation of the most costly component in GBS, the linear interferometer. 
In this section %will 
we introduce the high-level algorithmic optimization opportunity to overcome the challenge of infinite-dimensional qumode gate matrices and formulate our Bosonic compiler optimization problem.

%\subsection{High-Level Semantics of the Linear Interferometry}

%\TODO{High-level semantics of the linear interferometry}
%\textcolor{blue}{A typical gate in a Bosonic circuit has infinite dimension since the state of Bosonic mode lies in infinite-dimensional Hilbert space. Thus, it is impossible to optimize by considering the gate matrices.}

%\textcolor{blue}{In the GBS circuit, we are not interested in the matrix representation of the quantum gates. Instead, we focus on the parameters in each gate.} 

%\textcolor{blue}{
%Our compiler can directly analyze the GBS program upon the unitary of size $N\times N$ and decompose it into basic gates such as Phase Shifter and Beamsplitter by equation \eqref{eq1}. To optimize the circuit, we only need to tune the parameters of the gates and synthesize the circuit.}

\subsection{Opportunities from Decomposition Flexibility} \label{opp}
%\textcolor{blue}{A New Decomposition Method Using 2D Structure}}
%\TODO{Opportunities to approximate the unitary}

%The existing triangular decomposition method and rectangular decomposition method mentioned in Section \ref{Gaussian Boson Sampling} are based on the real device that has a one-dimensional structure, which means the two modes gate: Beamsplitter can only be added on the adjacent qumodes on the left and right.

The optimization opportunities come from the high-level algorithmic property of the $\mathbf{T}_{m,n}(\theta,\phi)$ transformations.
This transformation can be considered as using one entry in column $n$ to eliminate the entry in column $m$ in the unitary $\mathbf{U}$, similar to a Gaussian elimination process.
Overall, the decomposition of $\mathbf{U}$ is to find a series of $\mathbf{T}_{m,n}(\theta,\phi)$ transformations to convert the unitary $\mathbf{U}$ into a diagonal matrix.
Note that the original decomposition formula in Equation (\ref{eq1}) does not require any specific elimination order. Thus, the decomposition can be very flexible.
Existing decomposition methods~\cite{reck1994experimental,clements2016optimal} decompose the unitary into some special $\mathbf{T}_{m,n}(\theta,\phi)$'s where $n=m+1$ is always fixed. The optimization opportunities from the flexible decomposition were missed.

\begin{figure}[t]
    \centering
    \includegraphics[width=0.95\columnwidth]{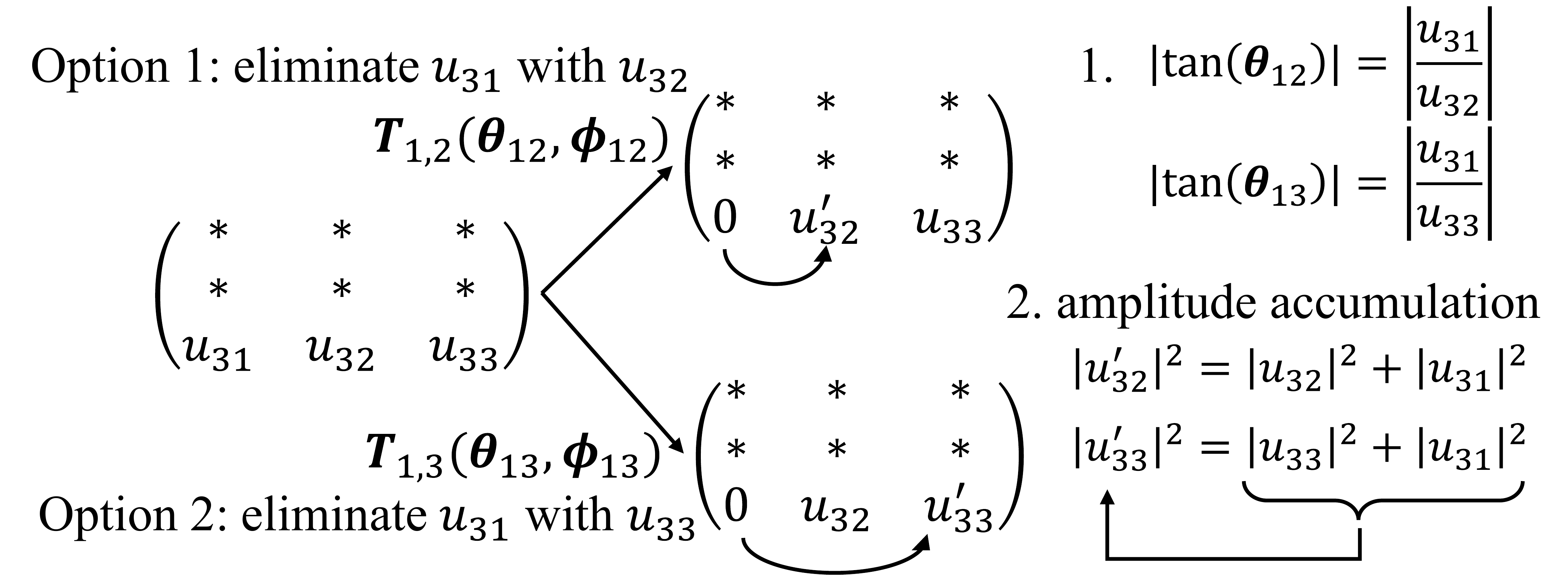}
    \vspace{-10pt}
    \caption{Example of flexible decomposition}
   \vspace{-10pt}
    \label{fig:elimination}
\end{figure}

We first illustrate the flexible decomposition of the linear interferometer unitary with the example in Fig.~\ref{fig:elimination}.
There are two decomposition options  in Fig.~\ref{fig:elimination} to perform elimination on the row $(u_{31}, u_{32}, u_{33})$.
The first option in the upper half of Fig.~\ref{fig:elimination} is to eliminate $u_{31}$ with $u_{32}$ using an MZI block $\mathbf{T}_{1,2}(\theta_{12},\phi_{12})$ on qumode 1 and 2. The parameters $\theta_{12}$ and $\phi_{12}$ are determined by $u_{31}$ and $u_{32}$ with following equation:
\begin{equation}
\begin{array}{ll}
     u_{31}e^{-i\phi_{12}}\cos(\theta_{12})-u_{32}\sin(\theta_{12}) = 0
     \label{eq2}
\end{array}
\end{equation}
We observe that there are two key properties of this elimination process that can help with the follow-up optimizations.

\textbf{First}, the rotation angle of the Beamsplitter in the generated MZI block satisfies:
%\begin{equation*}
 $  \vert \tan(\theta_{12}) \vert = \left\vert u_{31}/u_{32} \right\vert  $.
%\end{equation*}
And we can notice that the parameter $\theta_{12}$ will be very small when $\abs{u_{32}}$ is much larger than $\abs{u_{31}}$.
In this case, the Beamsplitter in the generated MZI block will be very close to an identity. % ~\ali{say something about how ``error'' (distance to identity) relates to beamsplitter angle, is it sin(theta) or something?}.
Theoretically, the distance between a Beamsplitter with a small rotation angle $\theta$ and the identity is bounded by $\sim\theta^2/N$.
%\jz{For a Beamsplitter with rotation angle $\theta$, its distance to identity will be upper bound by the value $2(1-\cos\theta)/N$ with N be the dimension by using the metric in Section~\ref{sec:setup}.}
%which means we use complex number $d$ to eliminate $c$. By comparing the real and imaginary parts, we need to solve the following equation:
%\begin{equation*}
%   \left\{\begin{array}{ll}
%   c_1\cos(\phi)+c_2\sin(\phi)=d_1\tan(\theta)\\
%   c_2\cos(\phi)-c_1\sin(\phi)=d_2\tan(\theta)
%   \end{array}
%   \right.
%\end{equation*}
%where we assume $c = c_1+ic_2$ and $d = d_1+id_2$. Notice that:
%\begin{equation*}
%   \vert \tan(\theta_{12}) \vert = \left\vert \frac{u_{31}}{u_{32}} \right\vert,  
%\end{equation*}
%the parameter $\theta_{12}$ will be very small when $\abs{u_{32}}$ is much larger than $\abs{u_{31}}$.

\textbf{Second}, all the matrices we used in decomposition are unitary, and the norm in every column and every row is always preserved. Therefore the amplitudes will be accumulated during the elimination. For example, in the first decomposition option in Fig.~\ref{fig:elimination}, the amplitude of the entry $u_{31}$ will be accumulated into the new amplitude $u_{32}^{\prime}$ after the elimination with $\abs{u_{32}^{\prime}}^2 = \abs{u_{31}}^2 + \abs{u_{32}}^2$.

Another option shown in Fig.~\ref{fig:elimination} is to eliminate $u_{31}$ with $u_{33}$.
This time the MZI block will be applied on qumode 1 and 3. The parameters and the amplitude accumulation will be different. This is a three-qumode example and the flexibility can be much larger with more qumodes.

\subsection{Compilation Problem Formulation}
%Here we formulate 
%\textbf{Hardware Characteristics:}
This paper also considers the hardware's characteristics.
Similar to qubit-based quantum hardware, Bosonic quantum hardware suffers from noise, and it is desirable to reduce the number of gates, especially the expensive two-qumode Beamsplitters.
We also consider the qumode connectivity constraints and only allow MZI blocks to be applied on physically-adjacent qumode pairs with native Beamsplitter support. %and  two-dimensional lattice coupling drived from recent experimental progress~\cite{} and schematic designs~\cite{}. 

Here we formulate the compilation problem of this paper. Given the input GBS program and the underlying hardware constraints, \myCompilerNameSpace needs to find 1) the logical-to-physical qumode mapping and 2) the decomposition of the linear interferometer unitary into hardware-supported MZI blocks.
Moreover, to mitigate the hardware noise effects, \myCompilerNameSpace aims to exploit the flexibility in decomposition such that the rotation angles in the generated Beamsplitters can be very small.
These Beamsplitters are very close to the identity and can thus be dropped to mitigate their high gate error without significantly affecting the overall GBS program. %\ali{
Importantly, our method does not merely approximate the program, but it does so at the point of decomposition and mapping, and we have full control over the compiler output's accuracy. 
%}

\section{Elimination Pattern Finding} \label{Zigzag Elimination Pattern Finding}

%\textcolor{blue}{With these two math properties in mind, we are now ready to design the new decomposition method.}

We will first introduce the unitary decomposition optimization in \myCompilerName. 
Our objective is to maximize the generation of MZI blocks with small Beamsplitter rotation angles.
Our decomposition approach will only generate MZI blocks that are compatible with the qumode without any remapping in the middle. 
We will assume a trivial logical-to-physical qumode mapping in this section and discuss the related further optimization in the next section.
Our design is based on the widely used two-dimensional lattice hardware coupling structure. 
But our idea and design flow can be generalized to other layouts like triangular or hexagonal arrays. %, hexagonal arrays, heavy-hexagon, etc.

%\textcolor{blue}{To mathematically decompose a general unitary need to recursively decompose the matrix from the last row to the first row, by pulling out one rotation at each time, finally the unitary can be written as Equation \ref{eq1}. } 

%\textcolor{blue}{When we want to execute the unitary on a real device by building basic rotation gates, we need to decide a way to put elements in unitary to real qumodes on the device. We call this process logical-to-physical mapping. }

%\textcolor{blue}{
%As a default, we will assume the entries in the first column of unitary are the first ones, which means they will be sent to the first qumode in the real device, other columns and qumodes are corresponding similarly, this map is called an identity map. The recursively decompose works as follows: from the last to the first, we put every row into the real device according to the identity map, then apply an elimination pattern to each row until there exists only one non-zero entry. Next, we will introduce this elimination pattern as well as its corresponding structure of real devices.}
% Figure~\ref{fig:unitarydec} shows a decomposition example of a 4-by-4 matrix.

% \begin{figure}[t]
%     \centering
%     \includegraphics[width=1.0\columnwidth]{fig/unitarydec.pdf}
%     \vspace{-15pt}
%     \caption{\textbf{General Unitary Decomposition}}
%     \vspace{-5pt}
%     \label{fig:unitarydec}
% \end{figure}

\subsection{Elimination Pattern Template}
\label{sec:eliminationpatterntemplate}
%\TODO{First summarize what are the design requirements.
%1. the pattern must be a connected graph
%2. maximize the large-small pairs (why we have the branches)
%3. since the amplitude is accumulated, we hope the accumulation stay in large entries (why we need a main path)}

%\textcolor{blue}{In This section we will construct a template for the ideal decomposition of the unitary, there are several requirements of the template to be achieved.}

We define an \textit{elimination pattern template} to represent how the entries in the linear interferometer unitary $\mathbf{U}$ are eliminated. 
In this template, each node represents a qumode.
When we eliminate one entry of one qumode $i$ using the entry of another qumode $j$, we use a directed edge to connect the two corresponding qumodes from $i$ to $j$.

\begin{figure}[t]
    \centering
    \includegraphics[width=1.0\columnwidth]{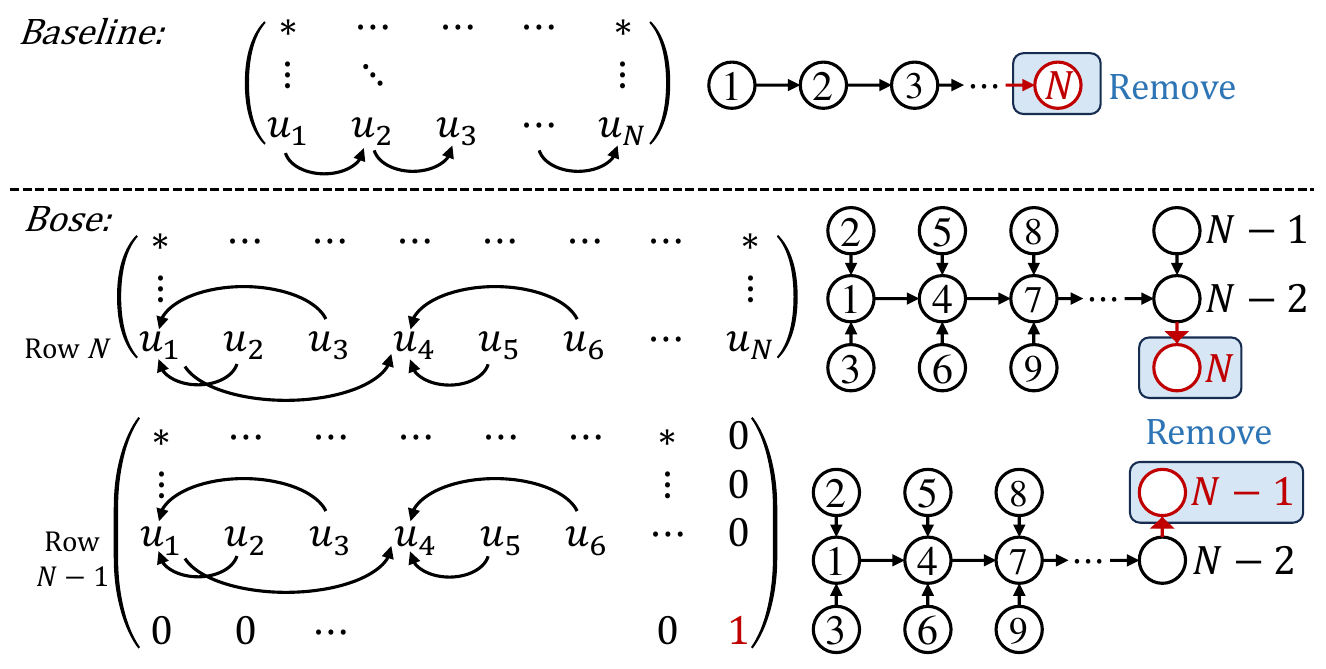}
    \vspace{-15pt}
    \caption{Elimination pattern template: baseline vs \myCompilerName}
    \vspace{-15pt}
    \label{fig:3byN}
\end{figure}

\textbf{Baseline elimination} For example, the template in the upper part of Fig.~\ref{fig:3byN} is the elimination template of existing decomposition methods~\cite{reck1994experimental,clements2016optimal}.
It has a chain structure.
$u_1$ is first eliminated with $u_2$ so there is an edge from qumode 1 to 2.
Then $u_2$ is eliminated with $u_3$, $u_4$ is eliminated with $u_3$, $\dots$, and finally $u_{N-1}$ is eliminated with $u_{N}$.
After the elimination of the last row finishes, $|u_N|$ should be 1 due to amplitude accumulation.
Then for the next row $N-1$, the last node $N$ in the elimination pattern is removed. 
The elimination of row $N-1$ will follow the same pattern from $u_1$ but terminate at $u_{N-1}$. 
The node $N-1$ is then removed and the elimination of row $N-2$ will begin.
Such process repeats from row $N$ to row $1$ and all the nodes in the elimination pattern are removed.

%Then $u_2$ is eliminated with $u_3$, and  finally $u_{}$

\myCompilerNameSpace leverages the elimination flexibility mentioned in Section~\ref{opp} and redesigns a new elimination pattern.
We first discuss the requirements and desired properties of such an elimination pattern graph.

First, the template must be a tree with all the directed edges from child nodes to parent nodes. As discussed in Section~\ref{opp}, the amplitudes of the entries are accumulated during the elimination process. The directed edges in the graph then naturally represent the flow of the amplitude accumulation.
To complete the elimination of one row in the unitary, all the amplitudes must be finally accumulated onto one entry and all other entries must be zeroed.
A tree with all the directed edges from child nodes to parent nodes can naturally represent how the amplitudes are accumulated from the leaf nodes to the root node.
The dependency is that the entry of a parent node can only eliminate the entry of one child node after the entry of this child node has eliminated the entries of all its child nodes.
For example, in the middle of Fig.~\ref{fig:3byN}, we must first eliminate qumode 2 and 3 with qumode 1 before we can further eliminate qumode 1 with qumode 4.

Second, recall that we hope to generate MZI blocks with small Beamsplitter rotation angles for further optimization and approximation.
This can be realized by eliminating a small entry with a large entry (recall the first property of elimination in Section~\ref{opp} and Fig.~\ref{fig:elimination}).
%, \jz{which is indicated by equation set 1 in Figure~\ref{fig:elimination}}. 
%\ali{(recall the $\tan{\theta} in Fig~\ref{fig:elimination}.)$}.
Meanwhile, the entries will become larger and larger as the elimination process gets close to the root due to the amplitude accumulation (the second property of elimination).
If we can attach small entries to these entries with accumulated large amplitudes, we can create more large-small eliminations. 
%there must exist some entries that will pass the large amplitudes finally to the root.
%thus we want to maximize the qumode pairs in the template that can hold large-small pairs in unitary. 

Finally, we also consider that this elimination pattern graph will later be physically realized in a two-dimensional lattice coupling structure. Although we have not considered the actual mapping onto hardware, we require that each node in this graph can have at most four neighboring nodes, so that it can be later mapped onto the hardware without introducing complicated transformations.

With all these considerations, our elimination pattern template graph design is shown in the lower part of Fig.~\ref{fig:3byN}.
Basically, we have a \textit{main path} in this elimination pattern reflecting the main flow of amplitude accumulation (the chain of nodes 1, 4, \dots). We can expect the amplitudes of previous nodes will be accumulated to the node throughout this main path, and they will be relatively large when they are used to eliminate other nodes.
Thus for each node in the main path, we attach some leaf nodes as the \textit{branches}. These branch nodes do not eliminate any other nodes so their amplitudes are expected to be small and large-small eliminations are created.
Considering that each node has at most four neighbors, we attach two leaf nodes to each main path node, except the first and last node in the main path.

%\textcolor{blue}{
%\jz{Change}
The elimination process is executed row by row from bottom to top. 
We start from row $N$ and eliminate $u_2$ and $u_3$ with $u_1$ first. 
Then $u_1$ is eliminated with $u_4$.
The elimination will follow the pattern on the right and finishes after accumulating all the amplitudes in row $N$ to $u_N$ and we have $|u_N| = 1$.
All other entries in the column $N$ will also become zero due to the column normalization of unitary matrices. 
Then we eliminate row $N-1$ and the node $N$ is removed from the elimination pattern since there is no amplitude in the last entry of row $N-1$.
We also need a small modification at the end of the pattern, flipping the edge direction from $N-1$ to $N-2$.
This will make the elimination terminate at the $N-1$ node, which is a leaf node in the tree, and then this node $N-1$ is removed when eliminating the next row $N-2$.
Such an elimination process repeats from row $N$ to $1$. 
The MZI blocks in the decomposed circuit can be generated using the parameters of the  $\mathbf{T}_{m,n}\left(\theta,\phi\right)$'s obtained in the elimination.

\textbf{Comparison against the baseline}
Bose has a new elimination pattern that can leverage the properties of flexible elimination.
By increasing the elimination happening between large and small entry pairs, our new elimination pattern can yield more MZI blocks with small rotation Beamsplitters to benefit follow-up optimization and approximation.

\subsection{Zigzag Elimination Pattern Embedding} \label{ZigzagEliminationPatternFinding}

\begin{figure}[t]
  \centering
  \begin{minipage}[t]{1\linewidth}
    \centering
    \includegraphics[width=0.8\columnwidth]{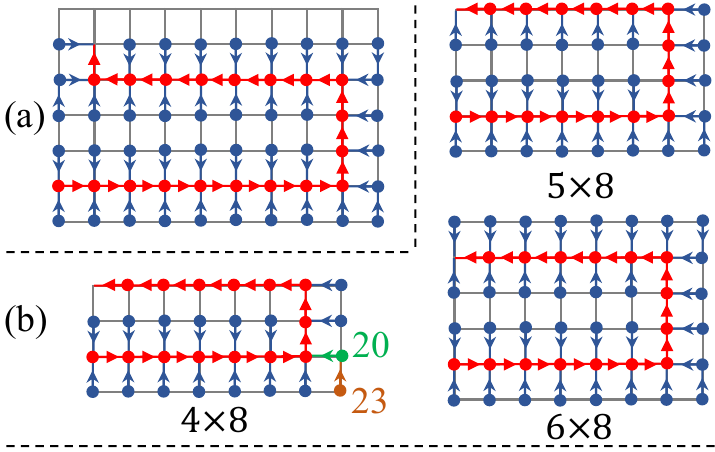}
    \vspace{-10pt}
    \label{fig:EmbeddingExample}
  \end{minipage}
  \begin{minipage}[t]{1\linewidth}
    \centering
    \includegraphics[width=0.85\columnwidth]{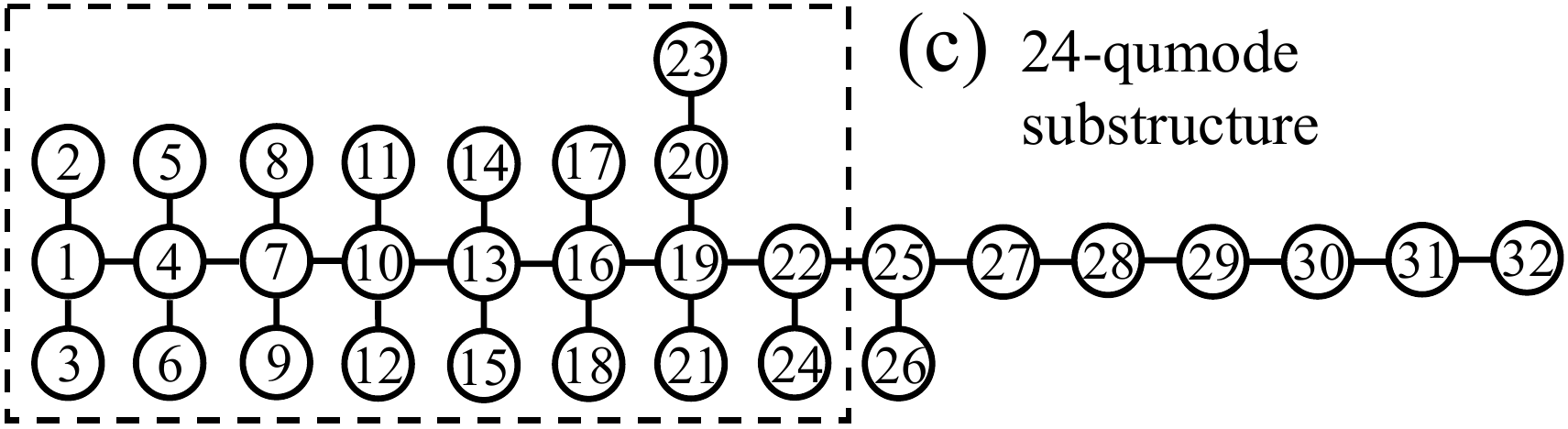}
    \vspace{-10pt}
    \caption{Example of elimination pattern embedding}
    \vspace{-15pt}
    \label{fig:decexample}
  \end{minipage}
\end{figure}

The next step is to embed the elimination pattern into the actual hardware coupling graph, a two-dimensional lattice.
We propose a Zigzag pattern embedding to maintain the overall structure in the original template. % with some branch nodes dropped.
%The strategy we adopted here is called \textbf{Zigzag Elimination Pattern Finding}, which means we drop out some of the ancillary modes in the template and embed the remaining modes into the rectangular device with a Zigzag shape.
A general procedure is shown in Fig.~\ref{fig:decexample} (a).
We start the embedding by aligning the main path with the longer edge of the two-dimensional lattice.
Here we start from the bottom left of the lattice and embed the pattern template in the first three rows.
The start node of the main path is denoted as the `start point'.
When the embedding reaches the right side, the main path will turn up and align with the column edge.
Around the turning point, we may be unable to attach two branch nodes to each main path node. % (node 23 in Figure~\ref{fig:decexample} (b)).
If one node cannot be directly connected to the main path, we can attach it to one branch node (e.g., nodes 23 and 20 in Fig.~\ref{fig:decexample} (b)).
After we reach the next three rows, our main path will turn to the left and follow similar patterns to add branches.
When we reach the left side, the main path will turn up again and finally formulate a Zigzag pattern.
Depending on the result of the number of rows modulo 3, we may need to drop some branch nodes at the end.
 Fig.~\ref{fig:decexample} (b) shows the three different cases.
Finally, the main path will end on either the left or right edge at a point denoted as the `end point'.

% \begin{figure}[t]
%     \centering
%     \includegraphics[width=1.0\columnwidth]{fig/real structure.pdf}
%     \vspace{-15pt}
%     \caption{\textbf{Embedding Example}}
%     \vspace{-5pt}
%     \label{fig:EmbeddingExample}
% \end{figure}

% \begin{figure}[t]
%     \centering
%     \includegraphics[width=1.0\columnwidth]{fig/dec example.pdf}
%     \vspace{-15pt}
%     \caption{\textbf{Example of Decomposition Pattern}}
%     \vspace{-5pt}
%     \label{fig:decexample}
% \end{figure}

\subsection{Sub-Pattern Selection}
The last step is to select some of the physical qumodes for the follow-up computation when the total number of qumodes on the device exceeds the total number of logical qumodes in the program. 
%We need another step to select the 
%The final step of the elimination pattern finding is the remove the redundant physical qumodes and only the number of physical qumodes that is equivalent to the number of logical qumodes in the program.
%In the above two subsections, we have figured out what kind of structure to adopt when we have a general rectangular device as well as how to use it to apply the unitary decomposition. This section will deal with the case when our total modes of the device exceed the total modes of the task we want to implement on GBS device. 
Recall that the purpose of our template pattern is to use the entries on the main path to eliminate entries on the branches to produce small rotation angles. 
Therefore, we will select the qumodes on the main path connecting to more branch qumodes and disregard those main path qumodes with fewer branch qumodes.

Our physical qumode selection will first label all the physical qumodes based on a breadth-first-search starting with the first qumode (`start point') in the template pattern. Fig.~\ref{fig:decexample} (c) shows an embedding structure that comes from a $4 \times 8$ device. 
As shown in Fig.~\ref{fig:decexample} (b), the qumodes that are far away from the start point will have fewer branches due to the edge of the two-dimensional lattice. 
We will choose the qumodes from the lower label to the higher label until the number of total qumodes is satisfied to implement the problem we are considering.
An example of selecting a 24-qumode substructure from a 32-qumode $4 \times 8$ device is in Fig.~\ref{fig:decexample} (c).

\section{Qumodes Mapping Optimization} \label{Qmodes Mapping}

%After we select a subset of physical qumodes with optimized decomposition, the next step is to determine the exact logical-to-physical qumode mapping.
In the previous section, we select a hardware-compatible elimination pattern based on the trivial mapping.
In this section, we will introduce how the logical-to-physical qumode mapping can be optimized to further improve the yield of small Beamsplitter rotation angles via permutation operations that come with no execution overhead.

\subsection{Motivating Example} \label{Mapping for Only One Row}
We introduce our qumode mapping algorithm with a motivating example considering the unitary decomposition in one row. 
Recall that the small rotation angles are expected to happen when we use a qumode in the main path of the pattern to eliminate another qumode on the branches. 
If we can map the qumodes with large entries in the unitary on the main path and those with small entries on the branches at the very beginning, the angles we produced during the elimination will be further reduced.
Suppose we have a 24-dimensional vector:
%\TODO{do not use the symbol a, it is too similar to the annihilation operator}
\begin{equation*}
   (l_{1}, l_{2}, l_{3}, \ldots, l_{24}) 
\end{equation*}
and we assume that their amplitudes satisfy $|l_{1}|\geq|l_{2}|\geq \ldots \geq|l_{24}|$ without loss of generality.
In this example, we will perform elimination on this row using the 24-qumode elimination pattern from Fig.~\ref{fig:decexample} (c).
%In this subsection, we consider the case that we only deal with row decomposition, the example is shown in Figure~\ref{fig:rowmappingexample} (a), and we assume $|a_1|>|a_2|> \ldots >|a_{11}|$.

%Our qumode mapping algorithm will optimize for this objective.
%As mentioned above, our objective is to place large entries in the main path nodes and small entries in the branch nodes.
%Therefore the generated MZI blocks will have even smaller beamsplitter rotation angles.
%Our main task here is to give a label to the device and map the bigger entry of the vector to the qumode in the device that has a lower label. The labeling process can be divided into labeling the main path and labeling the branches.
%The device example is taken from the previous section see Figure \ref{fig:decexample}, where we select 24-qumode structure from a 32-qumode $4 \times 8$ rectangular device.

 \begin{figure}[t]
    \centering
    \includegraphics[width=0.9\columnwidth]{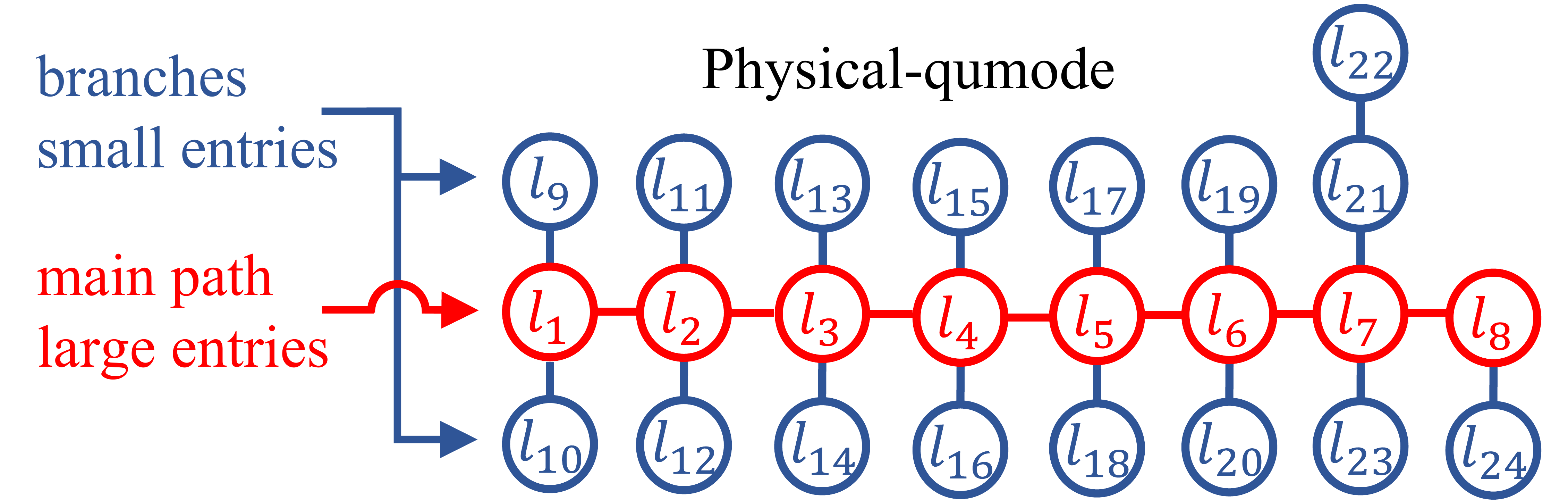}
    \vspace{-10pt}
    \caption{Example of logical-to-physical mapping}
    \vspace{-15pt}
    \label{fig:mapping}
\end{figure}

% The first step is to give lower label to the qumodes on the main path in the template from the start point, since we want to put large entry on the main path. As shown in Figure \ref{fig:mapping} \textcolor{blue}{(a)}, we give labels 0-7 to the main path qumodes.

% Next, we label the branches. The branches that are adjacent to the lower label qumode on the main path have the priority to get the lower label. As shown in Figure \ref{fig:mapping} (a), the ancillary qumode that is connected to qumode 0 get label 8 and 9. Also, two labels can be randomly sent to the branches when there exist two ancillary qumodes.

% A special case is that the qumode on the main path has 3 ancillary qumodes, like qumode 6 shown in \ref{fig:mapping} (a). We will first label the side that has two ancillary qumodes, the lower label is given to the qumode that is directly adjacent to the qumode on the main path. Then the side with one ancillary qumode will get the highest label among the three. 

A desired logical-to-physical qumode mapping of this example is depicted in Fig.~\ref{fig:mapping}.
The 8-qumode row in the middle is our main path, with its branches on two sides. 
Since we want large entries to appear in the main path, $l_{1}, l_{2}, \ldots, l_{8}$ will be mapped to the main path from the start point because they are the largest one. %, see Figure \ref{fig:mapping}.
After mapping the main path, we deal with the remaining branches, and the large entries remaining should be sent to the branch near the start point, thus $l_{9}, l_{10}$ are branches to $l_{1}$, other mappings are similar. 
In this way, the accumulated amplitude will be even larger when the elimination process along the main path is approaching the end point, because the branch nodes with large amplitudes are attached close to the start point.
The branch nodes attached to the main path nodes near the end point will have the smallest amplitudes.

A special case is that the qumode on the main path has 3 branch qumodes, like qumode contains $l_{7}$ shown in Fig.~\ref{fig:mapping}. We will first map the larger one to the long branch, and the smaller one to the short branch. In Fig.~\ref{fig:mapping}, we put $l_{21}$ and $l_{22}$ to long side and $l_{23}$ to the short side.

\subsection{Mapping via Permutation}
The qumode mapping of GBS in this paper is highly different from its counterpart in discrete-variable quantum computing, the qubit mapping problem. %, has been widely studied\TODO{~\cite{}} but qumode mapping of GBS in this paper is highly different.
The qubit mapping usually involves determining the initial logical-to-physical qubit layout and injecting SWAPs in the middle to resolve the dependencies for each two-qubit gate. 
%Our solution is based on the overall unitary rather than each individual gate.
In contrast, our mapping optimization will directly encode the mapping transition into the unitary, the high-level algorithmic representation of the linear interferometer. % leveraging the high-level .
%Therefore, the mapping transition can be directly encoded in the unitary.
We identify that the qumode mapping in GBS can be considered as adding permutation matrices before and after the unitary. 
And the permutation operations can be implemented without any additional gates. % and thus comes with no execution overhead. 
%by just relabeling the qumodes.
%Therefore, our qumode mapping optimization does not incur any additional qumode gates and thus comes with no execution overhead. %, even if the qumode SWAP gate exists.

%Before introducing the method for mapping a unitary matrix, we will prove the fact that doing the row permutation and column permutation to the unitary we want to encode in the device will not affect the final sampling result.
We first show that performing the row permutation and column permutation to the unitary encoded in the GBS interferometer will not affect the final sampling result.

\begin{figure}[t]
    \centering
    \includegraphics[width=0.9\columnwidth]{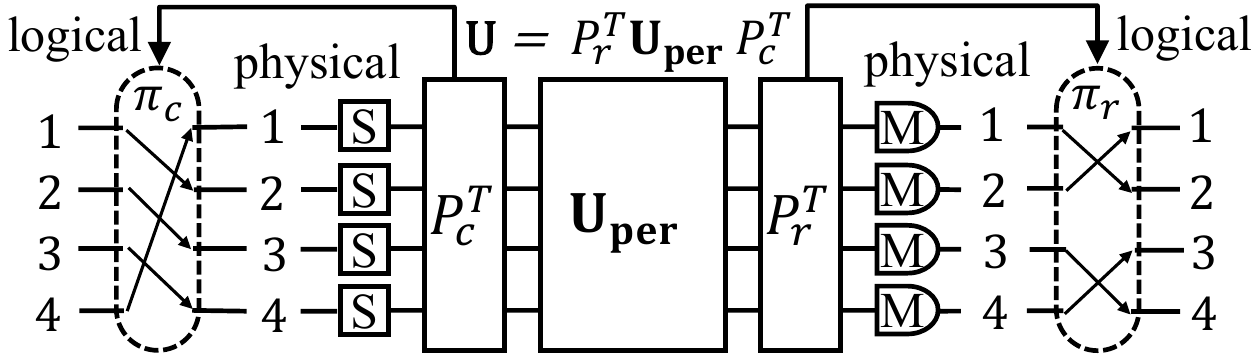}
    \vspace{-10pt}
    \caption{Mapping via Permutation}
    \vspace{-15pt}
    \label{fig:permutation}
\end{figure}

A matrix permutation can be expressed as:
\begin{equation*}
   \begin{array}{ll}
   \mathbf{U_{per}} = P_{r}\mathbf{U}P_{c} 
   \end{array}
\end{equation*}
where $\mathbf{U}$ is the original matrix, $\mathbf{U_{per}}$ is the matrix after permutation, $P_{r}$ and $P_{c}$ are row permutation and column permutation, respectively. We can rewrite the matrix $\mathbf{U}$ as:
\begin{equation*}
   \begin{array}{ll}
    \mathbf{U} = P_{r}^{T}\mathbf{U_{per}}P_{c}^{T} 
   \end{array}
\end{equation*}
In this case, if we want to encode unitary matrix $\mathbf{U}$ into the device, we can first encode the permutation $P_{c}^{T}$, then encode the unitary $\mathbf{U_{per}}$, and lastly encode another permutation $P_{r}^{T}$, as shown in Fig.~\ref{fig:permutation}.

%Usually in Bosonic devices, we use cavities to create qumodes, typically we will map qumode $i$ to the cavity $i$. 
Since the state preparation and measurement in GBS usually do not involve multi-qumode operations, $P_{c}^{T}$ can be done by changing the initial logical-to-physical qumode mapping. Suppose the $P_{c}^{T}$ is given by the following permutation:
\begin{equation*}
   \begin{array}{ll}
   i \to \pi_{c}(i)
   \end{array}
\end{equation*}
this relationship means we transfer the qumode $i$ into qumode $\pi_{c}(i)$, thus the physical qumode $\pi_{c}(i)$ after $P_{c}^{T}$ contains the qumode $i$ before this permutation. As a result, to implement the same transformation we can omit the permutation $P_{c}^{T}$ and map logical qumode $i$ into the physical qumode $\pi_{c}(i)$ directly, as shown on the left of Fig.~\ref{fig:permutation}.

%As for the permutation $P_{r}^{T}$, the case is the same. 
Similarly for the permutation $P_{r}^{T}$, suppose we get $output_{i}$ in the physical qumode $i$ after the unitary transformation $\mathbf{U_{per}}$, this output should lie in the physical qumode $\pi_{r}(i)$ if we execute permutation $P_{r}^{T}$, where $\pi_{r}(i)$ stands for the permutation given by $P_{r}^{T}$. Thus, instead of implementing the $P_{r}^{T}$ in the circuit, we can directly obtain the output of logical qumode $\pi_{r}(i)$ from the output of physical qumode $i$.

In summary, to obtain the GBS execution results with $\mathbf{U}$ as the linear interferometer from the GBS experiments using the permuted unitary $\mathbf{U_{per}}$, we just need to relabel the qumodes before and after the GBS program based on the permutation matrices  $P_{c}^{T}$ and  $P_{r}^{T}$. 

%The above discussion tells us that to get the same sampling results using only the permuted unitary $U_{per}$, we need to change the position of the qumodes we put in the cavity as well as the position of output information we get from the measurements.

\subsection{Two Properties of Elimination} \label{twoproperty}
\begin{figure}[t]
    \centering
    \includegraphics[width=0.6\columnwidth]{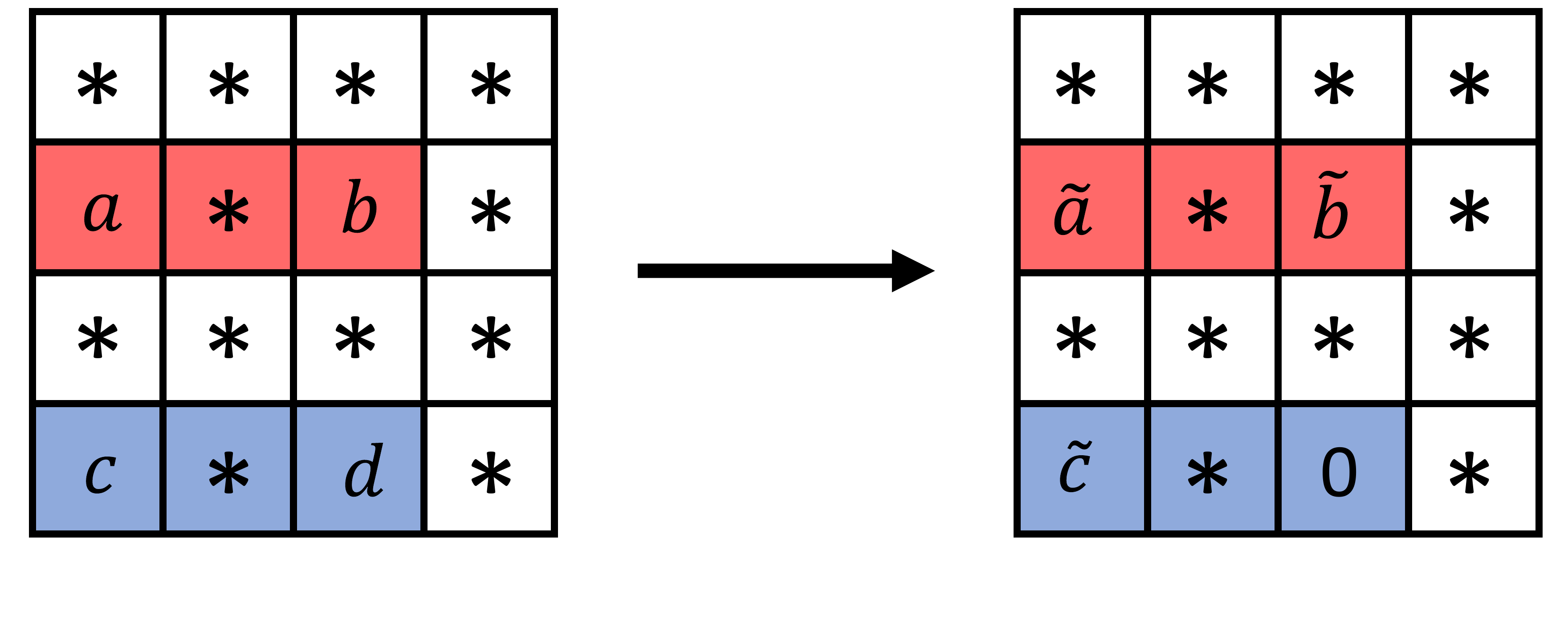}
    \vspace{-10pt}
    \caption{Property of elimination}
   \vspace{-15pt}
    \label{fig:property}
\end{figure}

%In this subsection, 
We first introduce two mathematical properties of the elimination process.
These two properties will guide the design of our qumode mapping algorithm.
%See Figure~\ref{fig:property} for example, 
We use the example in Fig.~\ref{fig:property}. Suppose $a$ and $c$ have large amplitudes while $b$ and $d$ are much smaller. We use $c$ to eliminate $d$ in the last row.

%The second property is that 
\textbf{First}, the elimination will not change the sum of the squares of the amplitudes in the region of a row containing all the entries that have changed in the elimination. 
In Fig.~\ref{fig:property}, we highlight the red region in the second row and the blue region in the last row. In the example, changes only happen in the first and third columns. The two highlighted regions contain all the entries that have changed values in the transformation. Note that:
\begin{equation*}
   \left\{\begin{array}{ll}
   \vert a \vert^2 + \vert b \vert^2 &= \vert \widetilde{a} \vert^2 + \vert \widetilde{b} \vert^2 \\
   \vert c \vert^2 + \vert d \vert^2 &= \vert \widetilde{c} \vert^2  
   \end{array}
   \right.
\end{equation*}
Thus the sum of the squares of the amplitudes in the highlighted region will not change.

This property allows us to %tells us that we can only 
focus on the amplitude in the region as a whole instead of each entry's specific amplitude in this region.
For example, in Fig.~\ref{fig:property}, when we decompose the last row, we may accumulate the amplitude of the first entry and second entry into the third entry. This process happens in the blue region, and then the third entry and the last entry can form a large-small pair to produce a small rotation angle. Because the sum of the squares of the amplitudes in the blue region is fixed, we can expect their amplitude accumulated as a whole to be large without checking each individual entry.
%we do not focus on which entry has a large amplitude, we only expect their amplitude accumulated as a whole to be large.

\textbf{Second}, the elimination for one row does not change the relative order of amplitudes of another row if the entries in both two rows are in decreasing order and the generated Beamsplitter rotation angle is small. %, which means 
That is, after the elimination, the amplitudes of $\widetilde{a}$ and $\widetilde{c}$ are still larger than that of $\widetilde{b}$. % remains small. % \review{I think you mean ``larger than''.}
This property can be understood with the formula that represents the elimination process:
\begin{equation*}
   \left\{\begin{array}{ll}
   \widetilde{a} = ae^{-i\phi}\cos(\theta)-b\sin(\theta), \ \widetilde{b} = ae^{-i\phi}\sin(\theta)+b\cos(\theta) \\
 %  \widetilde{b} &= ae^{-i\phi}\sin(\theta)+b\cos(\theta) \\
   \widetilde{c} = ce^{-i\phi}\cos(\theta)-d\sin(\theta), \   0 = ce^{-i\phi}\sin(\theta)+d\cos(\theta)
   %0 &= ce^{-i\phi}\sin(\theta)+d\cos(\theta)
   \end{array}
   \right.
\end{equation*}
The last equation gives the relationship that:
\begin{equation*}
   \vert \tan(\theta) \vert = \left\vert \frac{d}{c} \right\vert 
\end{equation*}
which indicates $\theta$ is small since we assume $d$ is small and $c$ is large. From the first equation, we have the following:
\begin{equation*}
\vert \frac{\widetilde{a}}{a} \vert \geq \vert \cos(\theta) \vert - \vert \frac{b}{a} \vert \vert \sin(\theta) \vert 
\end{equation*}
since $\theta$ and $\vert \frac{b}{a} \vert$ are small, the amplitude of $\widetilde{a}$ remains large. Similarly, the amplitude of $\widetilde{c}$ also remains large.

As for $\widetilde{b}$, we can derive the inequality that:
\begin{equation*}
\vert \frac{\widetilde{b}}{a} \vert \leq \vert \sin(\theta) \vert + \vert \frac{b}{a} \vert \vert \cos(\theta) \vert 
\end{equation*}
Since $\theta$ and $\vert \frac{b}{a} \vert$ are small, the amplitude of $\widetilde{b}$ remains small compared with $a$. 

This property is useful if we have a matrix in which the large entries appear in the beginning, and the small entries occur in the end for every row. Eliminating one of its rows won't change the order of absolute value in the remaining rows. As a result, if we find a good mapping for one of its rows (similar to the motivating example in Section~\ref{Mapping for Only One Row}), the elimination of other rows can still benefit from this mapping after the elimination of this row.

\subsection{Finding the Permutations}
%With the preparation in the last three subsections, we are now going to introduce the mapping method for the general unitary matrix.
We now describe our mapping method based on the properties and observations above. %The pseudo-code is in Algorithm~\ref{alg:permutationalgorithm}.
We explain it using the 24-qumode elimination pattern in Fig.~\ref{fig:mapping}. 

\begin{figure}[t]
    \centering
    \includegraphics[width=0.8\columnwidth]{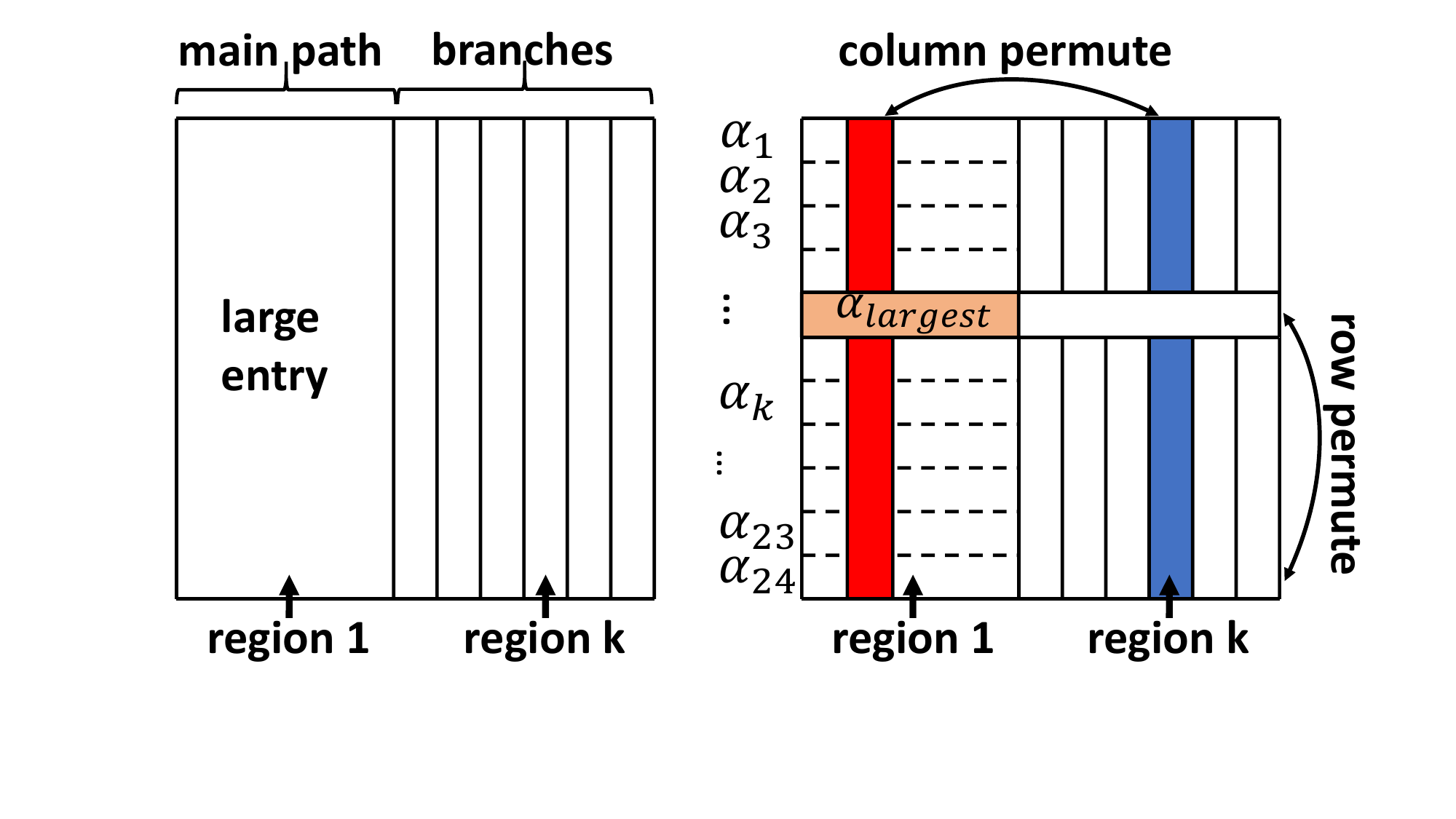}
    \vspace{-10pt}
    \caption{Column and row permutations}
    \vspace{-15pt}
    \label{fig:partition}
\end{figure}

In the elimination pattern, there are 8 qumodes on the main path and 16 qumodes on the branches.
Our objective is to map the large entries to the main path as much as possible.
The first property allows us to consider the amplitudes in a region with multiple columns. 
So our first step is to move large entries to the left side via the column permutations.
As depicted on the left of Fig.~\ref{fig:partition}, we vertically divide the unitary into multiple regions.
The first region is for the main path with 8 columns.
The following regions are for the branches and each region corresponds to one branch. 
They usually have one or two columns because the branches have one or two qumodes.

%All the rows of the unitary matrix can be divided into the same region according to the device structure. We still using Figure~\ref{fig:mapping} as an example. This structure can deal with problems containing 24 modes, the size of the unitary matrix is $24\times24$.

%The first region is the first 8 entries for the main path, the second region contains 1 entry for mode 8, the third region contains 1 entry for mode 9, and so on. The last three regions are 2 entries for modes 20 and 21, 1 entry for mode 22, and 1 entry for mode 23. See Figure~\ref{fig:partition} (a) for the matrix column partition. 

After the column partition, we will calculate the sum of the squares of the amplitudes in the main path region in each row as shown on the right of Fig.~\ref{fig:partition}. 
There summations are denoted as $\{\alpha_{1}, \ldots, \alpha_{24}\}$.
We denote the $K$-th largest one in this array as an indicator ($K$ can be around the size of half of the unitary dimension).
In practice, we select the value of $K$ that can generate more small Beamsplitter rotation angles ($\theta < 0.1$).% is less tha rotation angles. %  
This indicator generally represents the amplitudes of the main path region entries of the largest $K$ rows. % are. %are and we hope to
%, rearranging them in a decrease order we get $\{\alpha_{i_{1}}, \ldots, \alpha_{i_{24}}\}$. Then we will pick a comparison position $1\leq n\leq24$, we then choose $\alpha_{k_{n}}$ as the indicator of the original map, which tells us how the first n rows behave in region 1.
%Since we hope the qumodes mapped to the main path
%Since we want the indicator $\alpha_{k_{n}}$ as large as possible, we can exchange the column in region 1 with the column in other regions, see Figure~\ref{fig:partition} (b), we are exchanging the column j with column k. If the exchange leads to a larger $\alpha_{k_{n}}'$, we accept this exchange. 
Then we try to exchange the columns in the main path region 1 with the columns of other regions.
If we find that one exchange can increase the indicator, we will accept this exchange.
Such a process will merge the large entries to the main path and the first few branches.
With the recorded column exchanges, we will generate the overall column permutation.

%After seeking the best indicator $\alpha_{k_{n}}$ for region 1, we can move to other regions in order and do the same process until all the regions have been considered. 

We then find the row permutation. 
Since we decompose the unitary from the bottom row, we hope that rows with good numerical properties (i.e., those with large entries in the main path) are placed at the bottom.
With those good rows at the bottom, we can take advantage of the second property of the elimination process mentioned in Section~\ref{twoproperty}. These bottom rows are first executed and the elimination of these rows will not affect the good numerical properties of other rows.
%, and their uniform pattern can ensure the good pattern will not be disturbed by the decomposition of other rows.
Overall, our row permutation is generated by reordering the rows based on the sum of the squares of the amplitudes in the main path region of each row.

\section{Probabilistic Gate Dropout} \label{Unitary Approximation}

%In previous sections, we introduced new decomposition and mapping methods to help produce rotations with small angles. If we drop some of the small rotations in equation~\ref{eq1}, we can get an approximation unitary with high accuracy.
%\ali{conceptually why do we need this method if we are claiming that we can accurately predict approximation error? is it expensive to find the best angle dropouts that ``minimize'' error?}
The previous optimizations on linear interferometer unitary decomposition and qumode mapping have increased the occurrence of small rotation angles in the Beamsplitters.
In this section, we introduce the probabilistic gate dropout method that will select a sequence of rotation angles in the decomposition of unitary.
The purpose of this probabilistic dropout method is to drop those Beamsplitters with very small rotation angles with high probabilities while those with rotation angles near the threshold will be dropped more randomly to average over the algorithmic errors incurred by the approximation.
%by using the probability distribution to choose the rotations we want to take into account in unitary reconstruction. 

\textbf{Reconstructing High-Level Semantics}
One key advantage of \myCompilerNameSpace is that \myCompilerNameSpace can easily know the overall approximation effect after some gate dropout by reconstructing the high-level semantics from the decomposed gates. Recall the unitary decomposition formula in Equation~\ref{eq1}. 
Once we drop some Beamsplitters, we can reuse this formula to calculate the approximated unitary by setting the $\theta$'s in the corresponding MZI blocks to be 0.
For example, if we drop the Beamplitter in the second and the third MZI blocks. We can just use the following formula to obtain the approximated unitary $\mathbf{U_{app}}$ by setting $\theta_2 = \theta_3 = 0$:
%\begin{equation}%\footnotesize
   %\mathbf{U} = \Lambda\left(\mathbf{T}\left(\theta_1,\phi_1\right)\mathbf{T}\left(\theta_2,\phi_2\right)\mathbf{T}\left(\theta_{3},\phi_{3}\right)\cdots\right)
%\end{equation}
\begin{equation*}%\footnotesize
   \mathbf{U_{app}} = \Lambda\left(\mathbf{T}\left(\theta_1,\phi_1\right)\mathbf{T}\left(\theta_2=0,\phi_2\right)\mathbf{T}\left(\theta_3=0,\phi_{3}\right)\cdots\right)
\end{equation*}
Note that all the matrices in this formula have size $N\times N$ so that the overall approximated unitary can be calculated efficiently.
This allows us to easily know and tune how much approximation we will have during compilation time.

%The metric we use to measure the similarity of two matrices is:
%\begin{equation*}
%        accuracy = tr(A_{app}\cdot A^{T})/N
%\end{equation*}
%where A is an $N \times N$ matrix and $A_{app}$ is its approximation matrix.

%The method of picking rotations according to probability can be summarized below: 

We now introduce our gate dropout method which monitors the overall approximation during dropout gate selection. After the decomposition, we will have $N(N-1)/2$ MZI blocks with their Beamsplitter rotation angles $\{\theta_1, \theta_2, \ldots, \theta_{N(N-1)/2}\}$. 
These angles will be selected using the following procedure.
%\textbf{First,} we will select an accuracy threshold $\tau$. Then we find the angle threshold $\vert \Theta \vert$ such that if we omit the rotations in which the angel's absolute value is less than $\vert \Theta \vert$, we can get the approximation unitary whose accuracy is just above the accuracy threshold $\tau$. Suppose there are $M$ angles kept at this step.
%\textbf{Second, } all the angles are divided by $\vert \Theta \vert$. We will select a positive integer K, and raise the absolute values of the angles in the list to the $K$-th power. 
%$$\{\vert \frac{\theta_1}{\Theta} \vert^{K}, \vert \frac{\theta_2}{\Theta} \vert^{K}, \ldots, \vert \frac{\theta_{N(N-1)/2}}{\Theta} \vert^{K}\}$$
\begin{enumerate}
\item We will select an accuracy threshold $\tau$. Then we find the angle threshold $\vert \Theta \vert$ such that if we omit the rotations in which the angel's absolute value is less than $\vert \Theta \vert$, we can get the approximation unitary whose accuracy is just above the accuracy threshold $\tau$. Suppose there are $M$ angles kept at this step.
\item All the angles are divided by $\vert \Theta \vert$. We will select a positive integer $K$ and raise the absolute values of the angles in the list to the $K$-th power. 
$$\{\vert \theta_1/\Theta \vert^{K}, \vert \theta_2/\Theta \vert^{K}, \ldots, \vert \theta_{N(N-1)/2}/\Theta \vert^{K}\}$$
  %  \item Get the total $N(N-1)/2$ rotations angle by a decomposition method. Angels can be written as a list:
  %  \begin{equation*}
  %      angel \enspace list = [\theta_1, \theta_2, \ldots, \theta_{N(N-1)/2}]
 %   \end{equation*}
 %   \item Pick an accuracy threshold $\tau$. Find the angle threshold $\vert \Theta \vert$ such that if we omit the rotations in which the angel's absolute value is less than $\vert \Theta \vert$, we can get the approximation unitary whose accuracy is just over the accuracy threshold $\tau$. Also, record the number of angels whose absolute values exceed $\vert \Theta \vert$, and denote this number as M.
  %  \item Divide all the angels in the angel list by $\vert \Theta \vert$, pick a positive integer K, and raise the absolute value of the entry in the list to the Kth power.  
   % \begin{equation*}
    %    new \enspace list = [\vert \frac{\theta_1}{\Theta} \vert^{K}, \vert \frac{\theta_2}{\Theta} \vert^{K}, \ldots, \vert \frac{\theta_{N(N-1)/2}}{\Theta} \vert^{K}]
    %\end{equation*}
    \item We normalize these new angles to construct a probability distribution:
    \begin{equation*}
        p_i = \frac{\vert \theta_i/\Theta \vert^{K}}{\sum_{j = 1}^{N(N-1)/2} \vert \theta_j/\Theta \vert^{K} }
    \end{equation*}
    This distribution represents how likely an angle $\theta_i$ will be picked into matrix reconstruction. There are two special cases. If $K=1$, we are randomly sampling the Beamsplitters by their rotation angle amplitudes. If $K$ goes to infinite (usually 100 is enough), we simply drop the angles smaller than the angel threshold $\vert \Theta \vert$.
    \item We select $L$ as the number of iterations.  We select $M$ angles by the probability distribution for each iteration and reconstruct the approximation unitary matrix. We denote $\tau_{K}$ as the average fidelity of the $L$ iterations. %, calculate its accuracy. Take the average of the total L approximation unitary accuracy as a metric, we denote it as $\tau_{K}$. 
    \item We find the positive integer $K$ such that this process can maximize $\tau_{K}$. In this case, we are able to maximize the approximation accuracy with $M$ MZI blocks.
    %use a fixed amount of rotations, to be specific, M rotations, but the approximation accuracy is the highest we can get.
\end{enumerate}

After $M$, $\vert \Theta \vert$, and $K$ are determined, \myCompilerNameSpace will generate the GBS circuit for each sample.
One GBS program may require over thousands of repeated executions to obtain the final distribution.
%In the sampling process, we may take thousands of samples for a problem. 
In each execution, we will use the probability distribution above
%\begin{equation*%}
%        p_i = \frac{\vert \frac{\theta_i}{\Theta} \vert^{K}}{\sum_{k = 1}^{N(N-1)/2} \vert \frac{\theta_k}{\Theta} \vert^{K} }
%\end{equation*}
to select $M$ rotations and generate the  GBS circuit with their associated MZI blocks. 

%In this way, we can achieve a relatively high unitary approximation accuracy, while reducing the Beamsplitter we encountered in the circuit. As we use fewer gates in the circuit, our method is more resistant to photon loss than the classic triangular and rectangular circuit.

\section{Evaluation} \label{Evaluation}

%In this section, we utilize our optimization techniques to implement several Gaussian Boson Sampling tasks, the result shows significant advantages of our method under the photon loss.
In this section, we evaluate \myCompilerNameSpace by comparing with state-of-the-art baselines, analyze the effects of each optimization step, and perform end-to-end application performance studies.

\subsection{Experiments Setup}\label{sec:setup}

%Our experiments use the 'Gaussian' backend in the Strawberry fields.

%We choose four different benchmarks: dense subgraph, maximum clique, graph similarity, and vibration molecule. 

%For each benchmark, we also select a range of photon loss as well as a threshold of approximation accuracy for unitary. Finally, we execute the following four experiment settings:
\textbf{Experiment Configurations:} To illustrate the effect of each optimization step, we design four experiment configurations. 1. \textsf{`Baseline'} is to use the vanilla linear interferometer unitary decomposition~\cite{clements2016optimal} without any optimization. 2. \textsf{`Rot-Cut'} is to directly drop gates under the baseline decomposition to reach a given unitary approximation rate.
3. \textsf{`Decomp-Opt'} is to only use our optimized decomposition pattern with unitary approximation without qumode mapping optimization.
4. \textsf{`Full-Opt'} is to apply all \myCompilerNameSpace optimizations.

\begin{figure*}[t]
    \centering
    \vspace{3pt}
    \includegraphics[width=0.85\textwidth]{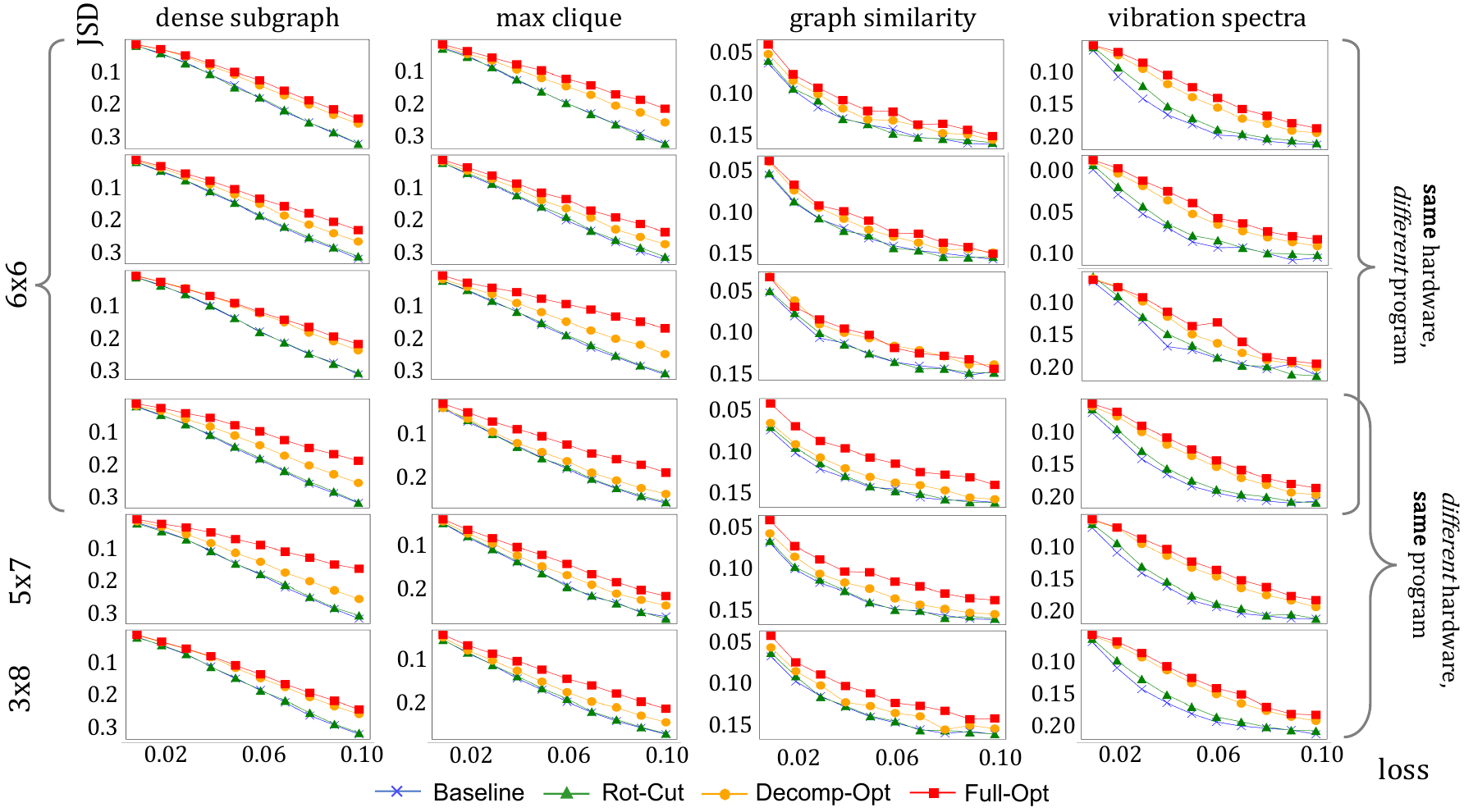}
    \vspace{-10pt}
    \caption{Overall GBS execution quality improvement with \myCompilerNameSpace optimizations} %ensen-Shannon Divergence\TODO{check, legend, label}}
    \vspace{-15pt}
    \label{fig:JSD}
\end{figure*}

\textbf{Hardware Configuration:} %This paper focuses on the bosonic superconducting quantum processor architectures. 
Similar to the qubit-based superconducting quantum architectures, two-dimensional lattice coupling is widely adopted in both recent experimental progress~\cite{wang2020efficient} and schematic design of superconducting Bosonic processors~\cite{blais2021circuit}.
We select three different 2D lattices: $6\times6$, $5\times7$, and $3\times8$.
%These architectures are derived 
%\TODO{find some experiment papers about }

\begin{table}[t]
  \centering
%   \vspace{-5pt}
  \caption{Benchmark Information}
  \vspace{-5pt} 
      \resizebox{\columnwidth}{!}{ 
    \begin{tabular}{|c|c|c|c|c|c|}
    \hline
    Benchmark & Qumode\# & Squeezing & Displacement & Phase Shifter & Beamsplitter  \\
   \hline
   DS & 24& 24 & 0 & 276 & 276\\
    \hline
   MC & 24&  24 & 0 & 276 & 276\\
    \hline
   GS & 24& 24 & 0 & 276 & 276\\
    \hline
   VS & 24& 24 & 24 & 276 & 276\\

    \hline
    \end{tabular}%
    }
    \vspace{-15pt}
  \label{tab:benchmark}%
\end{table}%
\iffalse
\begin{table}[t]
  \centering
%   \vspace{-5pt}
  \caption{Benchmark Information}
 % \vspace{-5pt} 
      \resizebox{\columnwidth}{!}{ 
    \begin{tabular}{|c|c|c|c|c|}
    \hline
    Benchmark &  Squeezing & Displacement & Phase Shifter & Beam Splitter  \\
   \hline
   DS &  24 & 0 & 276 & 276, 197\\
    \hline
   MC &  24 & 0 & 276 & 276, 210\\
    \hline
   GS &  24 & 0 & 276 & 276, 204\\
    \hline
   VS &  24 & 24 & 276 & 276, 167\\

    \hline
    \end{tabular}%
    }
    \vspace{-15pt}
  \label{tab:absgate}%
\end{table}%
\fi
\textbf{Benchmarks:} We select four different typical GBS applications, Dense Subgraph (DS), Maximum Clique (MC), Graph Similarity (GS), and Molecule Vibration Spectra Simulation (VS), with four program instances for each benchmark.
For DS, MC, and GS, we generate four random graphs of 24 nodes for each application and add edges between each pair of nodes with a probability of 0.7 to 0.9.
The numbers of different types of gates for the benchmarks are listed in Table~\ref{tab:benchmark}.
Note that for GBS programs, the numbers of gates mostly depend on the number of qumodes.
For VS, we select the molecule Pyrrole using the data from Strawberry Fields~\cite{killoran2019strawberry}. We simulate its vibrational spectra at four temperatures (1000K, 750K, 500K, and 250K).
All the programs in our benchmarks have 24 qumodes. 
This scale is limited by the classical simulator capability.
One 24-qumode GBS experiment simulation requires a few CPU hours, and we report the simulation results of over 1000 GBS experiments in this paper.

\textbf{Metrics:}
We use the Jensen-Shannon Divergence (JSD)~\cite{Jensen–S53:online} between the output distribution of the different experiment configurations and the standard output distribution as an application-independent metric to evaluate the improvement of \myCompilerName.
The standard distribution of each benchmark is generated by noise-free simulation of the original GBS program.
The compilation effect is indicated by the fidelity of the approximated unitary matrix of the linear interferometer and the number of gates. 
The fidelity is defined as $tr(\mathbf{U}_{app}\cdot \mathbf{U}^{\dagger})/N$ for $N$-qumode programs
%\end{equation*}
where $\mathbf{U}$ is the original $N \times N$ unitary and $\mathbf{U}_{app}$ is the approximated unitary.
We also adopt application-specific metrics at the end to provide a more intuitive understanding of the end-to-end benefit of \myCompilerName.

\textbf{Implementation:} We implemented \myCompilerNameSpace in Python and leveraged basic infrastructure in Strawberry Fields~\cite{killoran2019strawberry}. Our noisy GBS simulation experiments are executed on the `Gaussian' simulator backend in Strawberry Fields~\cite{killoran2019strawberry}.
To the best of our knowledge, this is already the most advanced simulator available which can allow us to accurately simulate 24-qumode GBS experiments with noise.
We simulate the gate photon loss error, which is the most significant error in Bosonic hardware and currently the only error type supported in the simulator, with an error rate ranging from 0.01 to 0.1 based on recent experimental data~\cite{lu2023high, chapman2023high}.
Each GBS experiment is sampled by 10000 times.
In the probabilistic gate dropout, the accuracy threshold $ \tau $ is set to be from $98.00\%$ to $99.96\%$, %\jz{for \textbf{\textsf{Rot-Cut}} as shown in table~\ref{tab:overall} (the threshold for \textbf{\textsf{Decomp-Opt}} and \textbf{\textsf{Full-Opt}} are even higher in each benchmark)}, 
then the Beamsplitter count $M$ and angle threshold $| \Theta |$  are determined as described in Section \ref{Unitary Approximation}. The power index $K$ is decided by repeating select $M$ Beamsplitters and calculating the average unitary approximation fidelity in 10000 samples, and its value ranges from 20 to 100. The experiments are executed on a server with 16 CPU cores and 128GB memory. %\TODO{add server configuration}

\begin{figure*}[t]
    \centering
    \includegraphics[width=0.85\textwidth]{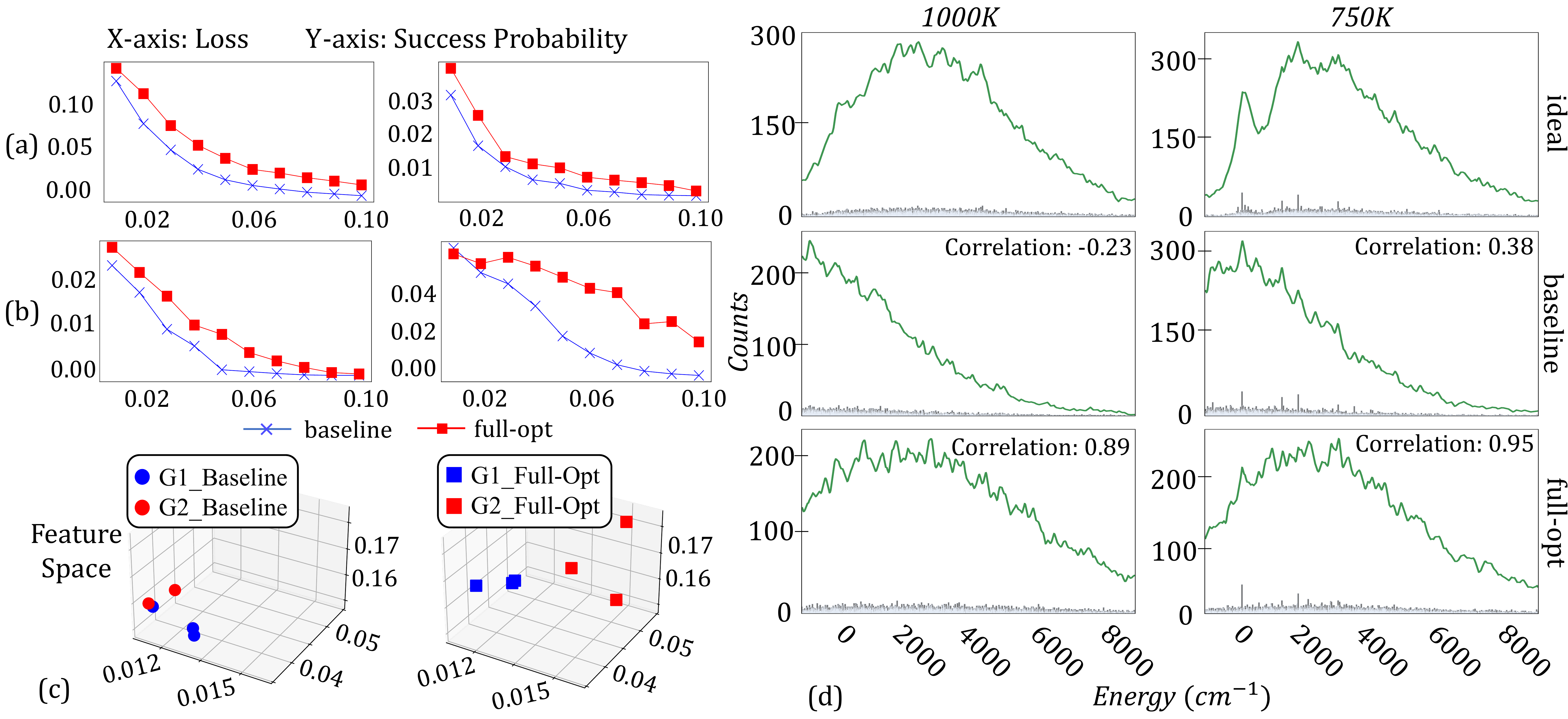}
    \vspace{-10pt}
    \caption{End-to-end performance improvement. (a) Dense Subgraph, (b) Maximum Clique, (c) Graph Similarity, (d) Vibration Spectra}
    \vspace{-15pt}
    \label{fig:end-to-end}
\end{figure*}

\begin{table}[t]
  \centering
%   \vspace{-5pt}
  \caption{Beamsplitter reduction and approximated unitary fidelity}
 % \vspace{-5pt} 
        \vspace{-5pt} 
      \resizebox{\columnwidth}{!}{ 
    \small\begin{tabular}{|c|c|c|c|}
    \hline
    Benchmark \& Fidelity &  Rot-Cut & Decomp-Opt & Full-Opt (Avg. Beamsplitter \#)  \\
   \hline
   DS, 99.90\% &  4.3\% & 16.8\% & 28.8\% (197)  \\
    \hline
  MC, 99.96\%&  5.0\% & 18.4\% & 24.1\% (210) \\
    \hline
   GS, 99.90\% &  3.4\% & 18.8\% & 26.0\% (204) \\
    \hline
   VS, 98.00\% &  11.8\% & 34.7\% & 39.6\% (167) \\

    \hline
    \end{tabular}%
    }
    \vspace{-15pt}
  \label{tab:overall}%
\end{table}%

\subsection{Overall Improvement}
\label{sec:overall}

%\jz{Gate reduction are refer to CNOT gate reduction.}

%In this section, we introduce the overall results on 6-by-6 device for the benchmarks, see Figure \ref{fig:JSD}. We use the Jensen-Shannon Divergence(JSD) as the matric, which can measure the distance between two probability distributions. Here, we calculate the JSD between the result from the loss-tolerant circuit and the circuit containing photon loss.  
We first apply \myCompilerNameSpace to approximate the linear interferometer to a certain  fidelity for all the four benchmarks on the $6\times6$ architecture, and the results are the first four rows in 
%Figure~\ref{fig:JSD} shows the improvement of \myCompilerNameSpace of all four benchmarks on the $6\times6$ architecture.
Fig.~\ref{fig:JSD}. % shows the overall improvement.
The X-axis is the loss rate and the Y-axis is the Jensen-Shannon Divergence (JSD) between the output distribution of the standard output and the distribution of the corresponding experimental configuration.
Each column represents one application and has four program instances for each.
Table~\ref{tab:overall} shows the fidelities of the approximated unitaries and the gate count reduction \myCompilerNameSpace is able to achieve.
A small JSD will indicate better performance.
It can be observed that as the loss increases, the JSD of \textsf{`Full-Opt'} grows much slower than that of \textsf{`Baseline'}.
On average, the JSD of \textsf{`Full-Opt'} can be reduced by  $31.6\%$, $33.8\%$, $12.6\%$, $26.4\%$, compared with \textsf{`Baseline'} for the DS, MC, GS, and VS benchmarks, respectively.
The great improvement comes from the fact that \myCompilerNameSpace can approximate the linear interferometer accurately using much fewer gates. 
As shown in Table~\ref{tab:overall}, \textsf{`Full-Opt'} can reduce  $28.8\%$, $24.1\%$, $26.0\%$, $39.6\%$ Beamsplitters but still maintain the fidelity of the linear interferometer unitary over %at \jz{more than} 
$99.90\%$, $99.96\%$, $99.90\%$, $98.00\%$, respectively.
The average remaining Beamsplitter count for \textsf{`Full-Opt'} is also in Table~\ref{tab:overall}. Note that the single-qumode gates are not changed in \myCompilerName.
In summary, with a tiny algorithmic error introduced in the unitary approximation, \myCompilerNameSpace can improve the overall performance by largely mitigating the hardware error.

\textbf{Importance of new decomposition and mapping:}
%The effect of each optimization step in \myCompilerNameSpace can be demonstrated by comparing the four experiment configurations.
%First, 
It can be observed that directly dropping the gates can hardly provide any improvement as the JSDs of \textsf{`Baseline'} and \textsf{Rot-Cut'} are very close (shown in Fig.~\ref{fig:JSD}).
The Beamsplitter reduction is also very small ($6.1\%$ on average in Table~\ref{tab:overall}), indicating a large hardware error.
These results suggest that the optimized decomposition pattern and qumode mapping is necessary to enable the optimizations in \myCompilerName.

\textbf{Effect of Each Step:}
The effect of the decomposition pattern optimization can be obtained by comparing \textsf{`Decomp-Opt'} against \textsf{`Baseline'}.
Fig.~\ref{fig:JSD} shows that \textsf{`Decomp-Opt'} can reduce the JSDs by $21.2\%$, $16.3\%$, $7.2\%$, $19.5\%$.
Table~\ref{tab:overall} shows that \textsf{`Decomp-Opt'} can reduce the gate count by $16.8\%$, $18.4\%$, $18.8\%$, $34.7\%$.
The effect of qumode mapping can be observed when comparing {\textsf{`Full-Opt}} and \textsf{`Decomp-Opt'}.
At the given approximated unitary fidelities, our logical-to-physical qumode mapping can further reduce the JSD  by $10.4\%$, $17.5\%$,  $5.4\%$, and $6.9\%$ on average for the four benchmarks. Table~\ref{tab:overall} also shows that the Beamsplitter count reduction is increased by $12.0\%$, $5.7\%$, $7.2\%$, $4.9\%$ ($7.4\%$ on average), respectively.
%and 
%Table~\ref{tab:overall} also shows that , our logical-to-physical qumode mapping can further increase the gate count reduction by $7.4\%$.  
The contribution breakdown between the decomposition optimization and the qumode mapping optimization is about $1.6:1$ and $3.0:1$ by comparing the JSD and Beamsplitter count reduction, respectively.
% \jz{We also report the average absolute gate count for benchmarks under \textbf{\textsf{Baseline}} and \textbf{\textsf{Full-Opt}} configuration in Table~\ref{tab:absgate}.}

%As shown in Figure \ref{fig:JSD}, the red line which represents the full optimization setting is strictly above the other three settings, its average distance reduction percentages are xx\%, xx\%, xx\%, xx\% for the four benchmarks respectively. Also, the yellow line which is only using the decomposition method shows its advantage over the baseline. The method that only cut rotations is similar to the baseline in our experiments.

\subsection{Hardware Structure Impact}
We also studied the impact of different hardware coupling structures, and the results are in the last three rows in Fig.~\ref{fig:JSD}. Each column represents one application, and each row represents one hardware structure.
We select one program instance for each application (one random graph for DS, MC, GS, and the 750K temperature simulation for VS). 
The results of other program instances of one application are similar.
It can be observed that the improvement of \myCompilerNameSpace is not affected by changing to a different 2D lattice structure.
On the $5\times7$ and $3\times8$ structures, the JSDs are reduced by $36.6\%$, $25.1\%$, $16.1\%$, $28.9\%$ for the four benchmarks, which is similar to the improvement on $6\times6$ structure. 

\subsection{End-to-End Application Performance Improvement}
%\ali{the results in this section are very good, highlight them in the abstract.}

%In this section, we introduce the direct metric and its result for the four benchmarks. Figure \ref{fig:end-to-end} shows the optimization results by directly measuring the outcome for the four tasks, for each task, four benchmarks are chosen.

In order to better understand the impact of \myCompilerNameSpace optimizations on the end-to-end application performance, we append the post-GBS data processing for all the benchmarks and evaluate them case-by-case. The details of the post-processing procedures for the selected benchmarks are out-of-scope, but they can be found in~\cite{bromley2020applications}. % and we develop them based on Strawberry Fields~\cite{killoran2019strawberry}.

\textbf{Dense Subgraph:} The GBS output will directly indicate a subgraph, and we measure the probability of successfully finding the densest subgraph of which the number of nodes is greater or equal to 10 in two of our random graphs, and the results are in Fig.~\ref{fig:end-to-end} (a).
The end-to-end success probability is increased by $64.1\%$ on average after \myCompilerNameSpace optimizations. % with \myCompilerName.

%Metric: the frequency that samples successfully find the densest subgraph in which nodes number greater or equal to 10. See Figure \ref{fig:end-to-end} (a) for the results.

\textbf{Maximum Clique:} The GBS output will serve as an initial trial in a follow-up clique finding subroutine. We measure the probability of successfully finding the clique whose nodes are greater or equal to 10 in two random graphs, and the results are in Fig.~\ref{fig:end-to-end} (b).
The end-to-end success probability is increased by $72.9\%$ on average after using \myCompilerName.

%Metric: the frequency that samples successfully find the clique in which nodes number greater or equal to 10. See Figure \ref{fig:end-to-end} (b) for the results.

\textbf{Graph Similarity:} The sampled output distribution will be converted into graph features. We randomly generate two highly-different graphs as two seeds and then generate two sets of similar graphs by adding small modifications to the two different seeds.
Fig.~\ref{fig:end-to-end} (c) shows the feature vectors in the feature space sampled from the graphs in the two similarity sets G1 and G2. 
On the left are the feature vectors obtained from {\textsf{`Baseline'}}. 
On the right are the feature vectors obtained with {\textsf{`Full-Opt'}}.
We can observe that the feature vectors are almost mixed for {\textsf{`Baseline'}} as the significant photon loss error tends to lose information about the sampled graphs.
But for the feature vectors of {\textsf{`Full-Opt'}}, they remain easily distinguishable for two clusters. 
We measure the distance between the two clusters' central positions in Fig.~\ref{fig:end-to-end} (c). It shows that the distance is increased by $135\%$ after using Bose.

%Metric: we create orbits and events according to the samples and make feature vectors. We place the original feature vectors of graphs as well as in the loss setting, the baseline experiment vectors and full-opt experiment vectors are chosen. See Figure \ref{fig:end-to-end} (c) for the results.

\textbf{Vibration Spectra:} We calculate the sampled molecule vibrational spectra using the GBS output. Fig.~\ref{fig:end-to-end} (d) shows the simulated vibrational spectra at 1000K and 750K with loss at 0.02.%\jz{In each sub-figure, 
The gray bars in the plot are the histogram of sampled energies and the green curve above is a Lorentzian broadening of the spectrum, which is a common practice in visualizing such a spectrum~\cite{bromley2020applications}. %}
The first row is the standard vibrational spectra.
The second row is the spectra obtained from the \textsf{`Baseline'} configuration.
The third row is the spectra obtained from {\textsf{`Full-Opt'}}.
Compared with {\textsf{`Baseline'}}, the results of {\textsf{`Full-Opt'}} are much more similar to the ideal spectra for both temperature settings.
Quantitatively, the Pearson correlation coefficients between the spectra generated by {\textsf{`Baseline'}} and the standard spectra are -0.23, 0.38 at 1000K and 750K, and the results of {\textsf{`Full-Opt'}} have much higher correlation coefficients 0.89 and 0.95. 
A correlation coefficient closer to 1 is better.
The spectra of the {\textsf{`Baseline'}} tend to shift to the low energy side due to the photon loss during the simulation. %\jz{The Pearson correlation coefficient between the spectra generated by \textbf{\textsf{Baseline}} and the standard spectra is -0.23, 0.38 under 1000K and 750K and 0.89 and 0.95 for \textbf{\textsf{Full-Opt}}, respectively. Since the closer the coefficient is to 1, the more similar the pattern we have, it quantitatively tells a much better performance of \myCompilerName.}
%\jz{explain the grey line}

%Metric: We calculate the Spectra according to our samples, and compare the Spectra graph directly. 

%See Figure \ref{fig:end-to-end} (d) for the results. The two columns stand for two temperature settings, the first row is the result that comes from the loss-tolerant circuit, and the next two rows are the baseline and full-opt circuit respectively under xx loss. We can clearly see that the Spectra is shifting toward zero when the loss is too big, but our optimization process can reduce the impact of photon loss.

\subsection{Scalability Study}
Although our GBS simulation is limited within 24 qumodes due to the complexity of classically simulating GBS, \myCompilerNameSpace comes with great scalability and can be applied to much larger GBS programs.
All the program analysis and compilation are performed on the high-level unitary representation of the linear interferometer, whose size grows linearly as the number of qumodes increases. 
The most time-consuming step is the matrix decomposition that has a complexity of $O(N^3)$ where $N$ is the number of qumodes.
Here we select seven numbers of qumodes from 10 to 500. % and generate a series of unitary matrices as the interferometer.
For each qumode count $N$, we randomly generate five unitaries as the interferometer, and then apply \myCompilerNameSpace to optimize it with a unitary approximation fidelity at $95\%$ on a $3\times \frac{N}{3}$ device. 
Table \ref{tab:scale} shows the average results, including the gate count reduction, the decomposition time, and the total time, of applying full optimization on five random unitaries.
Even for 500-qumode GBS programs, \myCompilerNameSpace can still reduce the number of gates by $27.1\%$ and the overall compilation time is less than half an hour.

%, and calculate the average performance results.
%In previous experiments, the dimension of our problem is 24. Here we generate random unitary from 8-40 dimension and apply our techniques to the original template. Table \ref{tab:scale} shows the average results of applying full optimization on five random unitaries, we can see that as the dimension goes up, we can use even less proportion of Beamsplitter to approximate the real one while maintaining the same accuracy.

\begin{table}[t]
  \centering
%   \vspace{-5pt}
  \caption{Performance at different problem scales (fidelity=0.95)}
  \vspace{-5pt}
      \resizebox{\columnwidth}{!}{ 
    \begin{tabular}{|c|c|c|c|c|c|c|c|c|c|}
    \hline
   Qumode \# & 10 & 15 & 20 & 60 & 100 & 200 & 500 \\
   \hline
    BS Gate \# drop & 29.33\% & 34.9\% & 31.6\% & 27.6\% & 27.5\% & 27.3\% & 27.1\%\\
    \hline
    Decomp time & 0.013s & 0.033s & 0.056s & 0.67s & 2.4s & 21.4s & 615.6s \\
    \hline
    Total time & 0.016s & 0.039s & 0.067s & 0.85s & 3.3s & 32.3s & 1071.1s \\
    \hline
    \end{tabular}%
    }
    \vspace{-15pt}
  \label{tab:scale}%
\end{table}%

\section{Related Work}
%\myCompilerName~is a compiler framework designed for near-term Bosonic quantum devices. 

\myCompilerName~is a compiler framework that can effectively and efficiently optimize (Gaussian) Boson sampling for Bosonic QC.
To the best of our knowledge, there is no closely related work targeting the approximation of linear interferometers or the qumode mapping problem.
We will briefly discuss the related works of the software frameworks for Bosonic QC, linear interferometer implementation, and approximated quantum compiler optimization.

\textbf{Bosonic Quantum Software Frameworks}
There have been several early efforts on the software framework for programming and compilation of Bosonic QC, such as the Xanadu's Strawberry Fields~\cite{killoran2019strawberry} and Quandela's Perceval~\cite{maring2023general} from industry, as well as the Bosonic Qiskit~\cite{stavenger2022c2qa} from academia. 
%Early efforts on programming and compilation for bosonic quantum computing include which provide basic programming interfaces for bosonic quantum computing, while optimizations/program transformations are mostly missing. \cite{reck1994experimental} and \cite{clements2016optimal} studied the basic approaches to decomposing linear interferometers into single- and two-qumode gates. %, which are the basic approaches to implement  
These works provide basic programming infrastructures for Bosonic QC but with almost no optimizations to the best of our knowledge.
In addition, \cite{kang2023leveraging} studied the low-level pulse compilation for individual qumode gates with analytical solutions for a single superconducting qubit-qumode pair. 
\cite{clement2022lov} designed a language to describe the linear optic quantum circuit. %, and they can normalize any polarization-preserving circuit into a unique triangular form.
\cite{rueda2021continuous} analyzed the continuous variable quantum compilation but failed to provide any actual optimizations. 
Unfortunately, none of them is able to simplify a Bosonic quantum program.
As an initial effort, this paper identifies several new optimization opportunities for Bosonic quantum compilation and proposes the corresponding effective optimization algorithms.

\textbf{Linear Interferometer Implementation}
To help design the linear optics experiments, previous works~\cite{reck1994experimental,clements2016optimal} have studied how to implement a linear interferometer with available optics instruments. 
Their solutions later serve as the linear interferometer implementation methods in Bosonic QC software frameworks like the Strawberry Fields~\cite{killoran2019strawberry} and Perceval~\cite{maring2023general}.
As introduced in Section~\ref{sec:eliminationpatterntemplate} (Fig.~\ref{fig:3byN}), they use a chain-structure elimination pattern with no remapping where it is hard to generate Beamsplitters with small rotations.
In contrast, this paper redesigns the elimination pattern and adaptively remaps the qumodes to yield more Beamsplitters with small rotations and thus benefit the follow-up optimization.
%can greatly reduce the number of MZI blocks but still maintain a high accuracy approximated implementation of a linear interferometer via a series of optimizations.
%\cite{reck1994experimental} and~\cite{clements2016optimal} can implement a linear interferometer precisely in two different (triangular and rectangular) yet fixed circuit shapes.
%All of these methods will yield quantum circuits containing exactly the same number of MZI blocks, which is $N(N-1)/2$ MZI blocks for $N$-qumode linear interferometers, and thus fail to reduce the execution cost.
%In contrast, this paper can greatly reduce the number of MZI blocks but still maintain a high accuracy approximated implementation of a linear interferometer via a series of optimizations.

\textbf{Approximated and Topology-Aware Quantum Compilation}
The approximated or topology-aware compilation has been widely explored in qubit-based quantum computing. 
Notable examples include a series of works~\cite{patel2022quest, davis2020towards,younis2021qfast,weiden2022wide,wu2021reoptimization} in the BQSkit project~\cite{younis2021berkeley} as well as some other works~\cite{madden2022best,martiel2022architectureaware, de2020architecture}.
%These works usually partition a quantum circuit into multiple small-scall blocks and then try to optimize them via topology-aware re-synthesis.
%The BQSKit project~\cite{younis2021berkeley} comes with a series of works to explore the approximated quantum circuit synthesis and optimization~\cite{patel2022quest, davis2020towards,younis2021qfast,weiden2022wide,wu2021reoptimization}.
%\cite{madden2022best} studied the approximated topology-aware quantum compilation for CNOT+rotation gate set. 
%\cite{de2020architecture},martiel2022architectureaware}
However, due to the difficulty in evaluating the gate matrices when the number of qubits is large, their approximation calculation and topology-aware resynthesis are usually limited to small-scale circuit blocks at each step or special types of circuits.
Moreover, such gate-matrix-based approaches are not applicable in Bosonic QC because of the built-in infinite-dimensional state spaces.
%The approximation 
% or some specific types of circuits.
This paper overcomes this challenge by leveraging the high-level scalable representation of general linear interferometers. And the proposed compilation techniques can easily control the overall approximation fidelity and perform large-scale topology-aware linear interferometer circuit synthesis and optimization.

\section{Conclusion}
This paper presented \myCompilerName, an innovative and efficient compiler optimization framework tailored for (Gaussian) Boson sampling on Bosonic quantum hardware. By addressing the challenge of infinite-dimensional qumode gate matrices and performing optimizations at a higher algorithmic level, \myCompilerName 
~significantly enhances the performance of Bosonic QC. The optimized qumode gate decomposition and logical-to-physical qumode mapping, along with the tunable probabilistic gate dropout method, demonstrate substantial improvements in reducing gate errors and approximating (Gaussian) Boson sampling with fewer gates. With \myCompilerName, the full potential of Bosonic QC can be effectively harnessed, hastening the onset of a wide range of practical applications that were previously computationally intractable.% and opening up new research opportunities. % The presented optimization framework marks a significant step forward in advancing Bosonic QC and opens up new opportunities for further research and development in quantum information processing.
%%%%%%% -- PAPER CONTENT ENDS -- %%%%%%%%
\section*{Acknowledgement}
We thank Nathan Wiebe and Steven Girvin for insightful discussions.
This work was partially supported by NSF CAREER Award 2338773, and the U.S. Department of Energy, Office of Science, National Quantum Information Science Research Centers, Co-design Center for Quantum Advantage (C2QA) under contract number DE-SC0012704.

%%%%%%%%% -- BIB STYLE AND FILE -- %%%%%%%%
\bibliographystyle{IEEEtranS}
\bibliography{refs}
%%%%%%%%%%%%%%%%%%%%%%%%%%%%%%%%%%%%

\end{document}